\pdfoutput=1
\RequirePackage{ifpdf}
\ifpdf 
\documentclass[pdftex]{sigma}
\else
\documentclass{sigma}
\fi

\begin{document}

\allowdisplaybreaks

\renewcommand{\PaperNumber}{073}

\FirstPageHeading

\ShortArticleName{Solutions Classif\/ication to the Extended Reduced Ostrovsky Equation}

\ArticleName{Solutions Classif\/ication\\ to the Extended Reduced
 Ostrovsky Equation}

\Author{Yury A. STEPANYANTS}

\AuthorNameForHeading{Y.A. Stepanyants}

\Address{Australian Nuclear Science and Technology,\\
Organisation PMB 1, Menai (Sydney), NSW, 2234, Australia}
\Email{\href{mailto:yuas50@gmail.com}{yuas50@gmail.com}}

\ArticleDates{Received July 14, 2008, in f\/inal form October
13, 2008; Published online October 26, 2008}

\Abstract{An alternative to the Parkes' approach [\href{http://dx.doi.org/10.3842/SIGMA.2008.053}{{\it SIGMA} {\bf 4} (2008), 053, 17~pages}] is
suggested for the solutions categorization to the extended reduced
Ostrovsky equation (the exROE in Parkes' terminology). The
approach is based on the application of the qualitative theory of
dif\/ferential equations which includes a mechanical analogy with
the point particle motion in a potential f\/ield, the phase plane
method, analysis of homoclinic trajectories and the like. Such an
approach is seemed more vivid and free of some restrictions
contained in [\href{http://dx.doi.org/10.3842/SIGMA.2008.053}{{\it SIGMA} {\bf 4} (2008), 053, 17~pages}].}


\Classification{35Q58; 35Q53; 35C05}

\section{Introduction}
\label{sec1} %
In paper \cite{Parkes} E.J.~Parkes presented a
categorization of solutions of the equation dubbed the extended
reduced Ostrovsky equation (exROE). The equation studied has the
form
\begin{gather}
\frac{\partial}{\partial x}\left({\cal D}^2u + \frac 12pu^2 +
\beta u \right) + q{\cal D}u = 0, \qquad \mbox{where} \quad {\cal
D} = \frac{\partial}{\partial t} + u\frac{\partial}{\partial x}
\label{eq1}
\end{gather}
with $p$, $q$, and $\beta$ being constant coef\/f\/icients. This
equation was derived from the Hirota--Satsuma-type shallow water
wave equation considered in \cite{Morrison-Parkes} (for details
see~\cite{Parkes}).

For stationary solutions, i.e.\ solutions in the form of
travelling waves depending only on one variable $\chi = x - Vt -
x_0$, this equation reduces to the simple third-order ODE:
\begin{gather}
\frac{d}{d\chi}\left[w\frac{d}{d\chi}\left(w\frac{dw}{d\chi}\right)
+ \frac 12pw^2 + (pV + \beta)w \right] + qw\frac{dw}{d\chi} = 0,
\label{eq2}
\end{gather}
where $V$ stands for the wave speed and $w = u - V$. (Note, that
in many contemporary papers including \cite{Parkes} authors call
such solutions simply ``travelling-wave solutions''. Such
terminology seems not good as nonstationary propagating waves also
are travelling waves. The term ``stationary waves'' widely used
earlier seems more adequate for the waves considered here.) In
paper \cite{Parkes}, equation~\eqref{eq2} was reduced by means of a
series of transformations of dependent and independent variables
to an auxiliary equation whose solutions were actually categorized
subject to some restrictions on the equation coef\/f\/icients, viz.:
\begin{gather*}
p + q \ne 0, \qquad qV -\beta \ne 0 
\end{gather*}
(one more restriction on the constant of integration for that
auxiliary equation, $B = 0$,  was used in \cite{Parkes}). Under
these restrictions, solutions to equation~\eqref{eq2} were found in
analytical form and corresponding wave prof\/iles were illustrated
graphically. Among solutions obtained there are both periodic and
solitary type solutions including multivalued loop periodic waves
and loop-solitons.

Similar loop solutions to exROE and some other equations were
earlier obtained by Ji-Bin Li~\cite{Li} who came to the conclusion
that loop solutions actually are compound solutions which consist
of three dif\/ferent independent branches. These branches may be
used in various combinations representing several types of
stationary propagating singular waves (waves with inf\/inite
gra\-dients). This conclusion completely coincides with the
conclusion of paper \cite{Stepanyants} where a complete
classif\/ication of stationary solutions of ROE was presented. ROE
derived by L.A.~Ostrov\-sky~\cite{Ostrovsky} in 1978 as a model for
the description of long waves in a rotating ocean (see~\cite{Stepanyants} and references therein) can be treated as a
particular case of exROE with $p = q$ and $\beta = 0$ (see~\cite{Parkes}).

\looseness=1
Below an analysis of stationary solutions to equation~\eqref{eq2} is
presented by the direct method avoiding any redundant
transformations of variables. The method used is based on the
phase plane concept and analogy of the equation studied with the
Newtonian equation for the point particle in a potential f\/ield.
Such approach seems more vivid and free of aforementioned
restrictions. This work can be considered also as complementary to
paper~\cite{Parkes} as the analysis presented may be helpful in
the understanding of basic properties of stationary solutions of
equation~\eqref{eq2}.

\section[Mechanical analogy, potential function and phase-plane method]{Mechanical analogy, potential function\\ and phase-plane method}
\label{sec2}

Equation \eqref{eq2} can be integrated once resulting in
\begin{gather}
w\frac{d}{d\chi}\left(w\frac{dw}{d\chi}\right) + \frac 12(p +
q)w^2 + (pV + \beta)w  = C_1, \label{eq4}
\end{gather}
where $C_1$ is a constant of integration. By multiplying this
equation by $dw/d\chi$ and integrating once again, the equation
can be reduced to the form of energy conservation for a point
particle of unit mass moving in the potential f\/ield $P(w)$:
\begin{gather}
\frac 12 \left(\frac{dw}{d\chi}\right)^2 + P(w) = E, \label{eq5}
\end{gather}
where the ef\/fective ``potential energy'' as a function of
``displacement'' $w$ is
\begin{gather}
P(w) = \frac{p + q}{6}w - \frac{C_1}{w} - \frac{C_2}{w^2},
\label{eq6}
\end{gather}
and $C_2$ is another constant of integration. The constant $E =
-(pV + \beta)/2$ plays a role of the total energy of the particle,
i.e.\ the sum of the ``kinetic energy'', $K =
(1/2)(dw/d\chi)^2$, and the ``potential energy'', $P(w)$. As
follows from equation~\eqref{eq5}, real solutions can exist only for $E
\ge P(w)$. Various cases of the potential function \eqref{eq6} are
considered below and corresponding bounded solutions are
constructed. Unbounded solutions are not considered in this paper
as they are less interesting from the physical point of view;
nevertheless, their qualitative behavior becomes clear from the
general view of corresponding phase portraits.

\section[Particular case: $p + q = 0$]{Particular case: $\boldsymbol{p + q = 0}$}
\label{sec3} %

Consider f\/irst a particular case when the coef\/f\/icients in
equation~\eqref{eq1} are such that $p + q = 0$. Note, this is one of
the cases which were omitted from the consideration in paper~\cite{Parkes}. The potential function~\eqref{eq6} simplif\/ies
in this case. However, a variety of subcases can be distinguished
nevertheless even in this case depending on the coef\/f\/icients $C_1$
and $C_2$. All these subcases are studied in detail below.

\begin{figure}[t]
\centerline{\includegraphics[width=150mm]{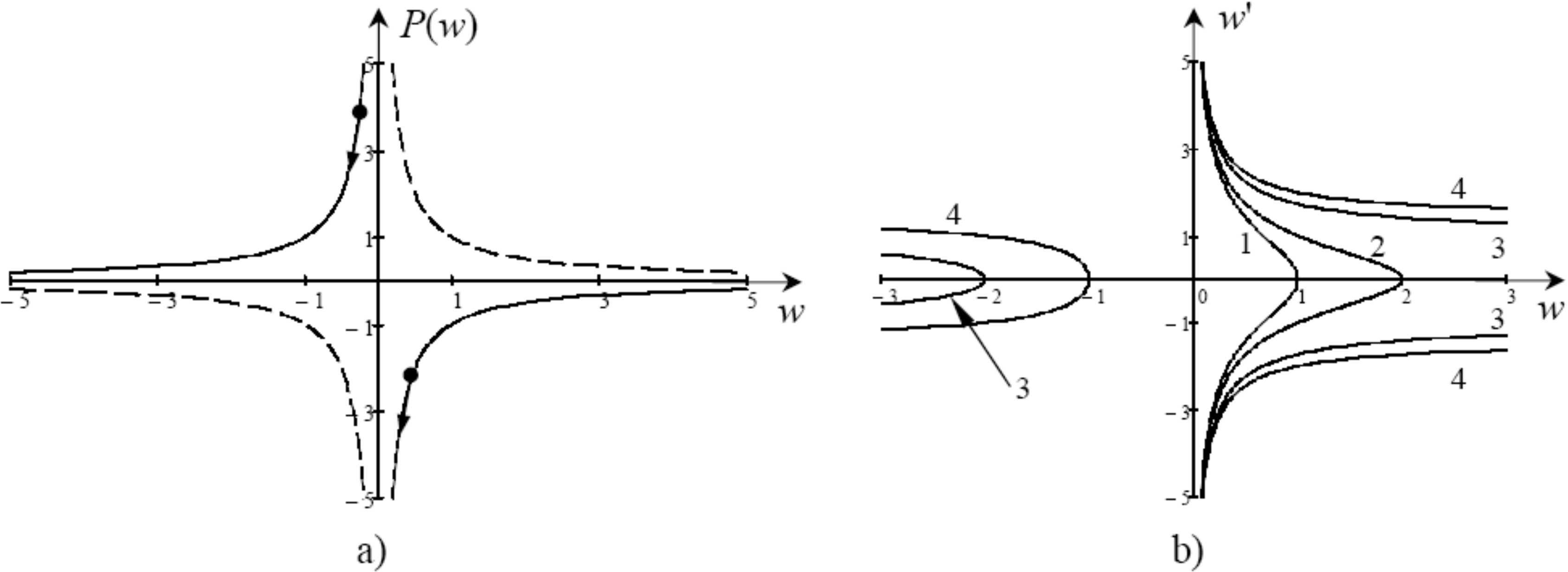}}
\caption{a) Potential function for the case $p + q = 0$, $C_2 = 0$
and two values of $C_1$: $C_1 = 1$ (solid lines), and $C_1 = -1$
(dashed lines). Two dots illustrate possible motions of point
particles in the potential f\/ield. b) Phase plane corresponding to
the potential function with $C_1 = 1$ and dif\/ferent values of $E$.
Line 1: $E = -1$; line 2: $E = -0.5$; lines 3: $E = 0.5$; lines 4:
$E = 1$.}
\label{f01}%
\end{figure}

{\bf 3a.} If $C_2 = 0$, the potential function represents
a set of antisymmetric hyperbolas located either in the f\/irst and
third quadrants or in the second and fourth quadrants as shown in
Fig.~\ref{f01}a. The corresponding phase plane $(w, w')$, where
$w' = dw/d\chi$, is shown in Fig.~\ref{f01}b for $C_1 = 1$ (for
$C_1 = -1$ the phase plane is mirror symmetrical with respect to
the vertical axis). For other values of $C_1$ phase portraits are
qualitatively similar to that shown in Fig.~\ref{f01} for $C_1 =
1$.

Analysis of the phase portrait shows that there are no bounded
solutions for any positive $E$; corresponding trajectories both in
the left half and right half of the phase plane go to inf\/inity on~$w$ (see, e.g., lines~3 and~4 in Fig.~\ref{f01}b).
Meanwhile, solutions bounded on $w$ do exist for negative values
of $E$ (i.e.\ for $V > - \beta/p$), but they possess inf\/inite
derivatives when $w = 0$. Consider, for instance, motion of an
af\/f\/ix along the line~2 in Fig.~\ref{f01}b ($C_1 = 1$) from $w' =
\infty$ towards the axis $w$ where $w' = 0$. The qualitative
character of the motion becomes clear if we interpret it in terms
of ``particle coordinate'' $w$ and ``particle velocity''~$w'$
treating~$\xi$ as the time. The motion originates at some ``time''
$\xi_0$ with inf\/inite derivative and zero ``particle coordinate''
$w = 0$. Then, the ``particle coordinate'' $w$ increases to some
maximum value $w_{\max} = -C_1/E$ ($E < 0$) as the ``particle
velocity'' is positive. Eventually it comes to the rest having
zero derivative $w' = 0$ and $w = w_{\max}$. Another independent
branch of solution for the same value of $E$ corresponds to the
af\/f\/ix motion along the line 1 from the previously described rest
point at axis $w$ towards $w' = -\infty$ and $w = 0$.

\begin{figure}[t]
\centerline{\includegraphics[height=55mm]{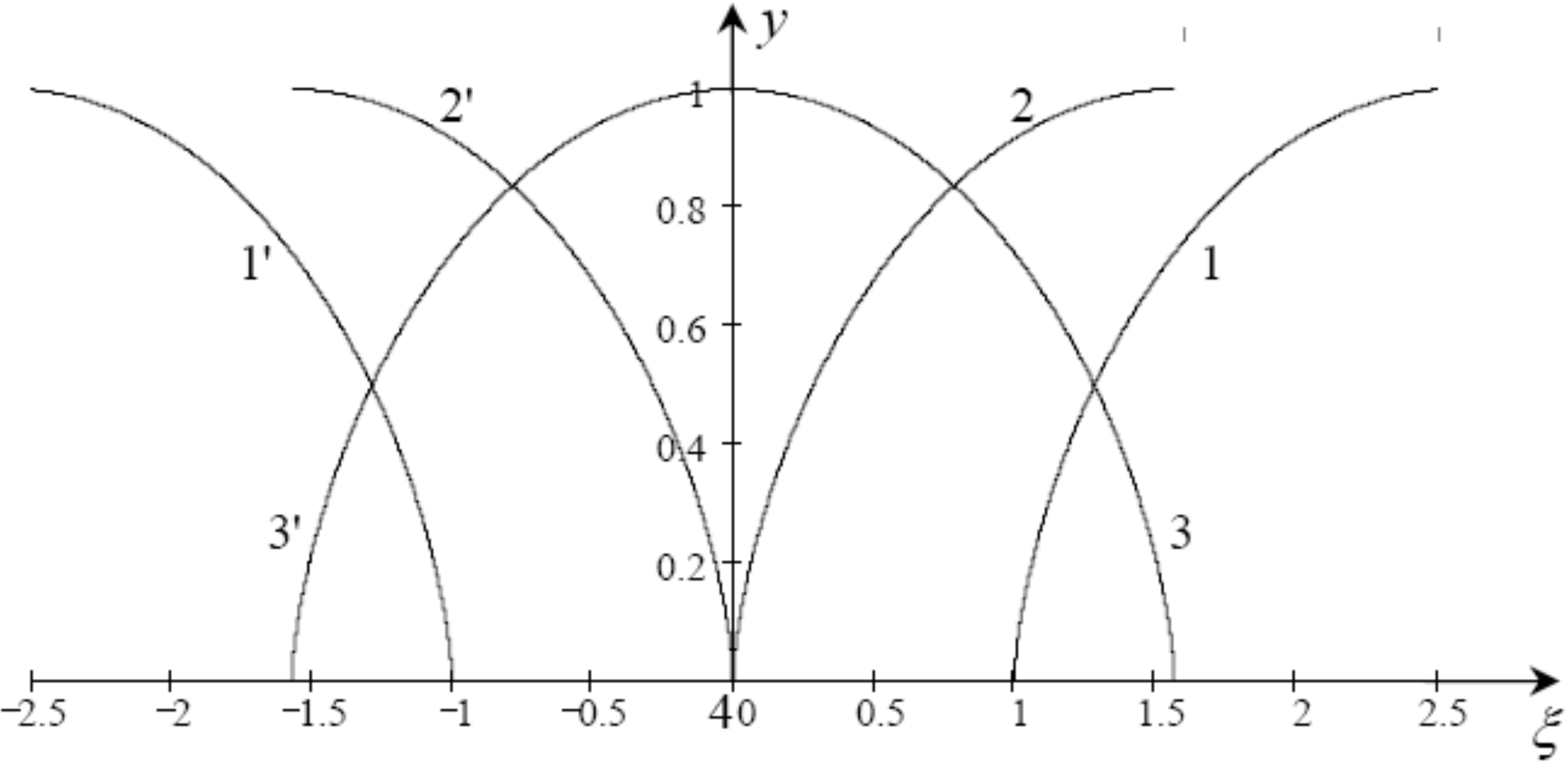}}
\caption{Various particular solutions described by
equation~\eqref{eq7}.}
\label{f02}%
\end{figure}

All bounded analytical solutions for this case can be presented in
the universal implicit form:
\begin{gather}
\xi(y) - \xi_0 = \pm\left[\arctan{\left(\sqrt{\frac{y}{1 -
y}}\right) - \sqrt{y(1 - y)}}\right], \label{eq7}
\end{gather}
where $y = -Ew/C_1$, $\xi = -\sqrt{2}(-E)^{3/2}\chi/C_1$  and
$\xi_0$ is an arbitrary constant of integration. This solution
consists of two independent branches which correspond to signs
plus or minus in front of the square brackets in equation~\eqref{eq7}.
Each branch is def\/ined only on a compact support of axis $\xi$:
either on $-\pi/2 \le \xi - \xi_0 \le 0$ or on $0 \le \xi - \xi_0
\le \pi/2$ (see lines~1 and~$1'$ in Fig.~\ref{f02}). With the
appropriate choice of constants  $\xi_0$ one can create a variety
of dif\/ferent solutions, e.g., the $V$-shape wave (see lines~2
and~$2'$), or a smooth-crest compacton, i.e.\ a compound
solitary wave def\/ined only for $|\xi - \xi_0| \le \pi/2$ (see
lines 3 and $3'$). Using a translational invariance of solutions
and their independency of each other, one can create periodic or
even chaotic sequences of compactons randomly located on axis
$\xi$.

\begin{figure}[t]
\centerline{\includegraphics[height=60mm]{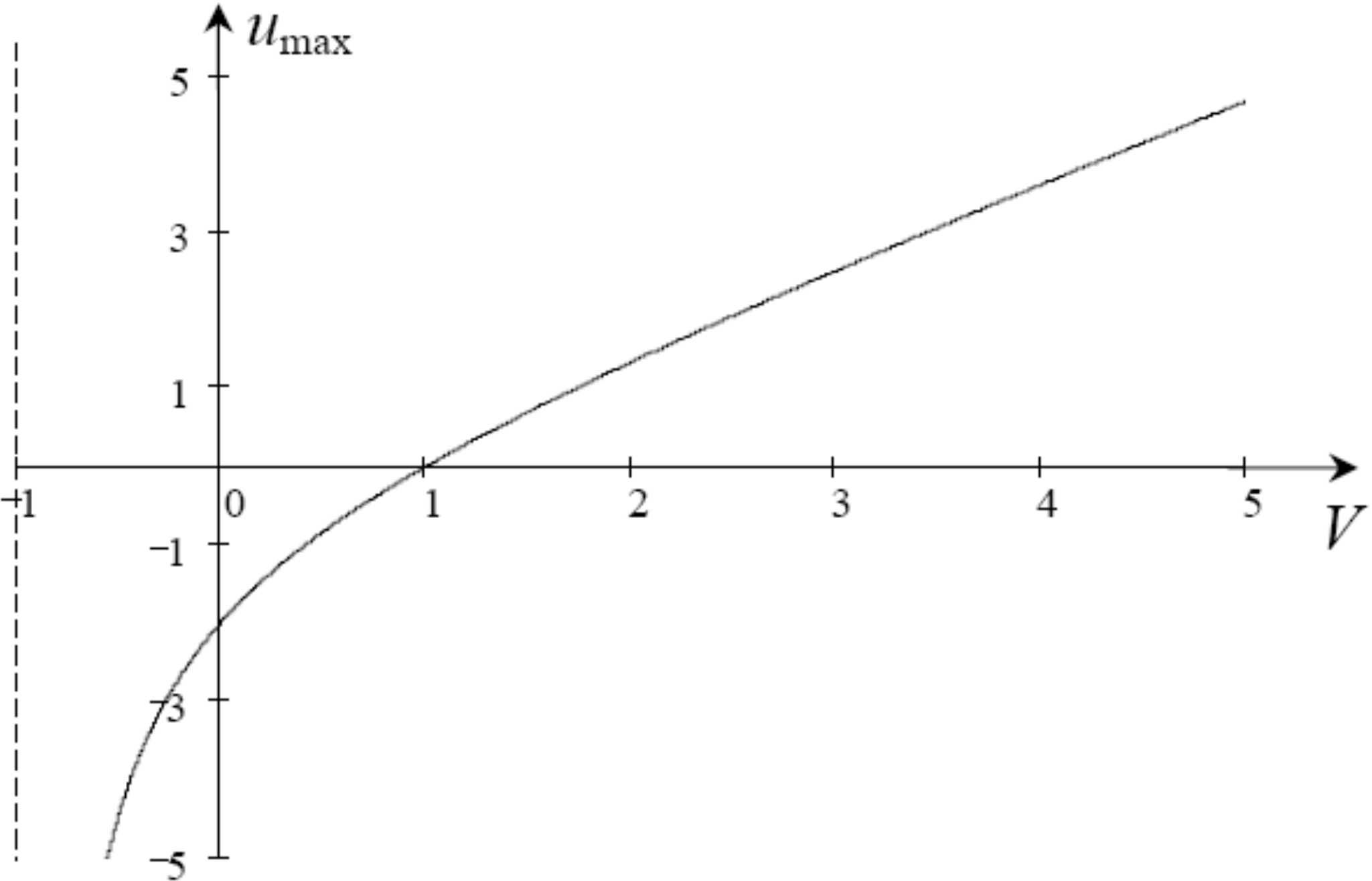}}
\caption{Maximum of the compacton solution~\eqref{eq7} against
speed in the original variables, equation~\eqref{eq8}. Dashed vertical
line corresponds to the limiting value of $V = -\beta/p$. The plot
is generated for $C_1 = p = \beta = 1$.}
\label{f03}%
\end{figure}

The maximum of the function $y(\xi)$, $y_{\max} = 1$, corresponds
in terms of $w$ to $w_{\max} = -C_1/E$. Using the relationship
between $w$ and the original variable $u$ (see above), as well as
the def\/inition of the constant $E$, one can deduce the
relationship between the wave extreme value (wave maximum) and its
speed:
\begin{gather}
u_{\max} = V - \frac{C_1}{E} = V + \frac{2C_1}{pV + \beta}.
\label{eq8}
\end{gather}
Taking into account that we consider the case of $C_1 = 1$, and
negative values of $E$ are possible only when $V > -\beta/p$, the
plot of $u_{\max}(V)$ is such as presented in Fig.~\ref{f03}.

As follows from equation~\eqref{eq8}, a wave is entirely negative
($u_{\max} < 0$), when
\[
V < \frac{1}{2p}\left(\sqrt{\beta^2 -
8pC_1} - \beta\right),
\]
provided that $p < \beta^2/(8C_1)$. At
greater values of $V$, the wave prof\/ile contains both positive and
negative pieces, and for certain value of $V$ the total wave
``mass'' $I = \int u(\chi)d\chi$ vanishes (the integral here is
taken over the entire domain where function $u(\chi)$ is def\/ined).

\begin{figure}[t]
\centerline{\includegraphics[width=150mm]{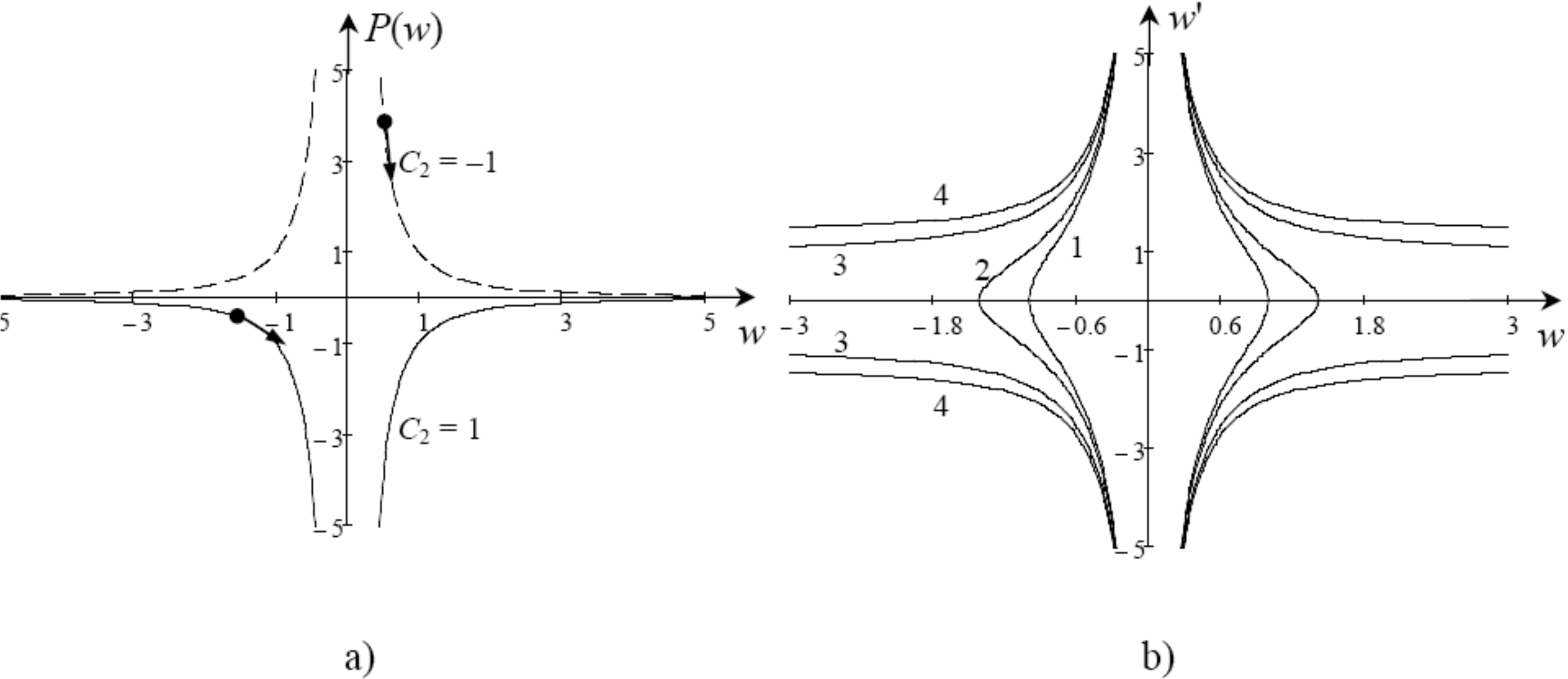}}

\caption{a)~Potential function for the case $p + q = 0$, $C_1 = 0$
and two values of $C_2$: $C_2 = 1$ (solid lines), and $C_2 = -1$
(dashed lines). Two dots illustrate possible motions of point
particles in the potential f\/ield. b)~Phase plane corresponding to
the potential function with $C_2 = 1$ and dif\/ferent values of $E$.
Line~1: $E = -1$; line~2: $E = -0.5$; lines~3: $E = 0.5$; lines~4:
$E = 1$. All lines are symmetrical with respect to axis $w'$ and
are labelled only in the left half of the phase plane.}
\label{f04}
\end{figure}

{\bf 3b.} A similar analysis can be carried out for the
case when $C_1 = 0$, $C_2 \ne 0$. The potential function in this
case represents a set of symmetric quadratic hyperbolas located
either in the f\/irst and second quadrants or in the third and
fourth quadrants as shown in Fig.~\ref{f04}a for $C_2 = \pm 1$.
The corresponding phase plane is shown in Fig.~\ref{f04}b for $C_2
= 1$ only (there are no bounded solutions for $C_2 = -1$,
therefore this case is not considered here). For other positive
values of $C_2$ phase portraits are qualitatively similar to that
shown in Fig.~\ref{f04}b.

\looseness=1
Analysis of the phase portrait shows that there are no bounded
solutions for $C_2 = -1$, as well as for $C_2 = 1$ and any
positive $E$ (see, e.g., lines~3 and~4 in Fig.~\ref{f04}b);
they exist however for $C_2 = 1$ and negative values of $E$, but
possess inf\/inite derivatives at some values of $\chi$. In
normalized variables $y = (-E/C_2)^{1/2}w$, $\xi =
-E(2/C_2)^{1/2}\chi$ all possible solutions can be presented in
terms of independently chosen function branches describing a unit
circle in one of the four quad\-rants,~i.e.
\begin{gather}
(\xi - \xi_0)^2 + y^2 = 1, \label{eq9}
\end{gather}
where $\xi_0$ is an arbitrary constant of integration.

\begin{figure}[t]
\centerline{\includegraphics[height=75mm]{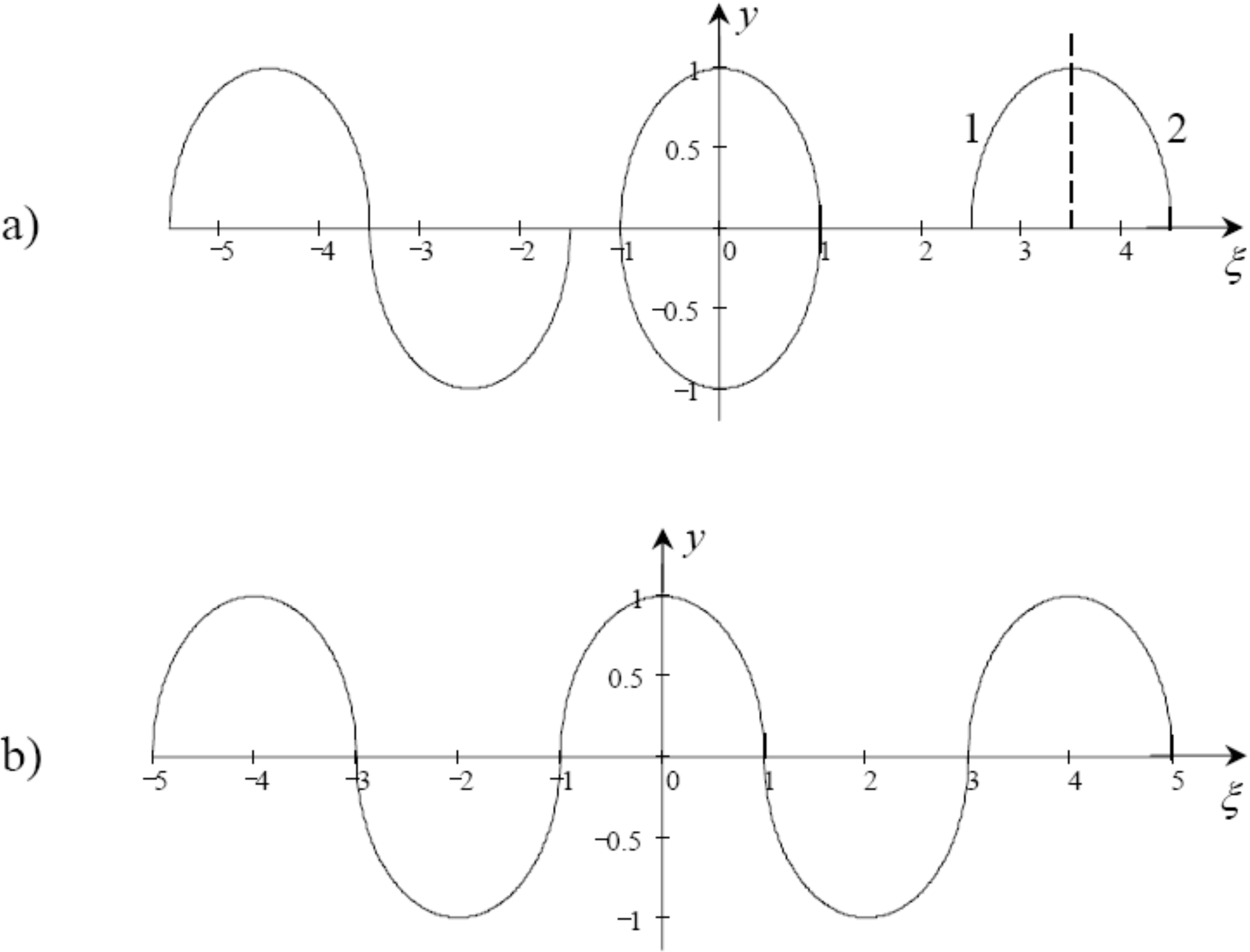}}

\caption{a) Some examples of pulse-type waves described by
equation~\eqref{eq9}: $N$-shaped wave; circle wave and semicircle
compacton. b)~One of the examples of a periodic wave with inf\/inite
derivatives at $y = 0$, $\xi = 2n + 1$, where $n$ is an entire
number.}
\label{f05}%
\end{figure}

Playing with the constant $\xi_0$ one can create again a variety
of compacton-type solutions including multi-valued solutions. Some
examples of solitary compacton solutions are shown in
Fig.~\ref{f05}a; they include $N$-shaped waves, multi-valued
circle-shaped waves and semicircle positive-polarity pulses (due
to symmetry, the polarity of the f\/irst and last waves can be
inverted). In addition to those, various periodic and even chaotic
compound waves can be easily constructed; one of the possible
examples of a periodic solution is shown in Fig.~\ref{f05}b. Each
positive or negative half-period of any wave consists of two
independent branches originating at $y = 0$ and ending at $y = \pm
1$. The same is true for the pulse-type solutions shown in
Fig.~\ref{f05}a; they consist of independent symmetrical branches
as shown, for example, for the semicircle pulse in Fig.~\ref{f05}a
where they are labelled by symbols 1 and 2.

\begin{figure}[t]
\centerline{\includegraphics[height=55mm]{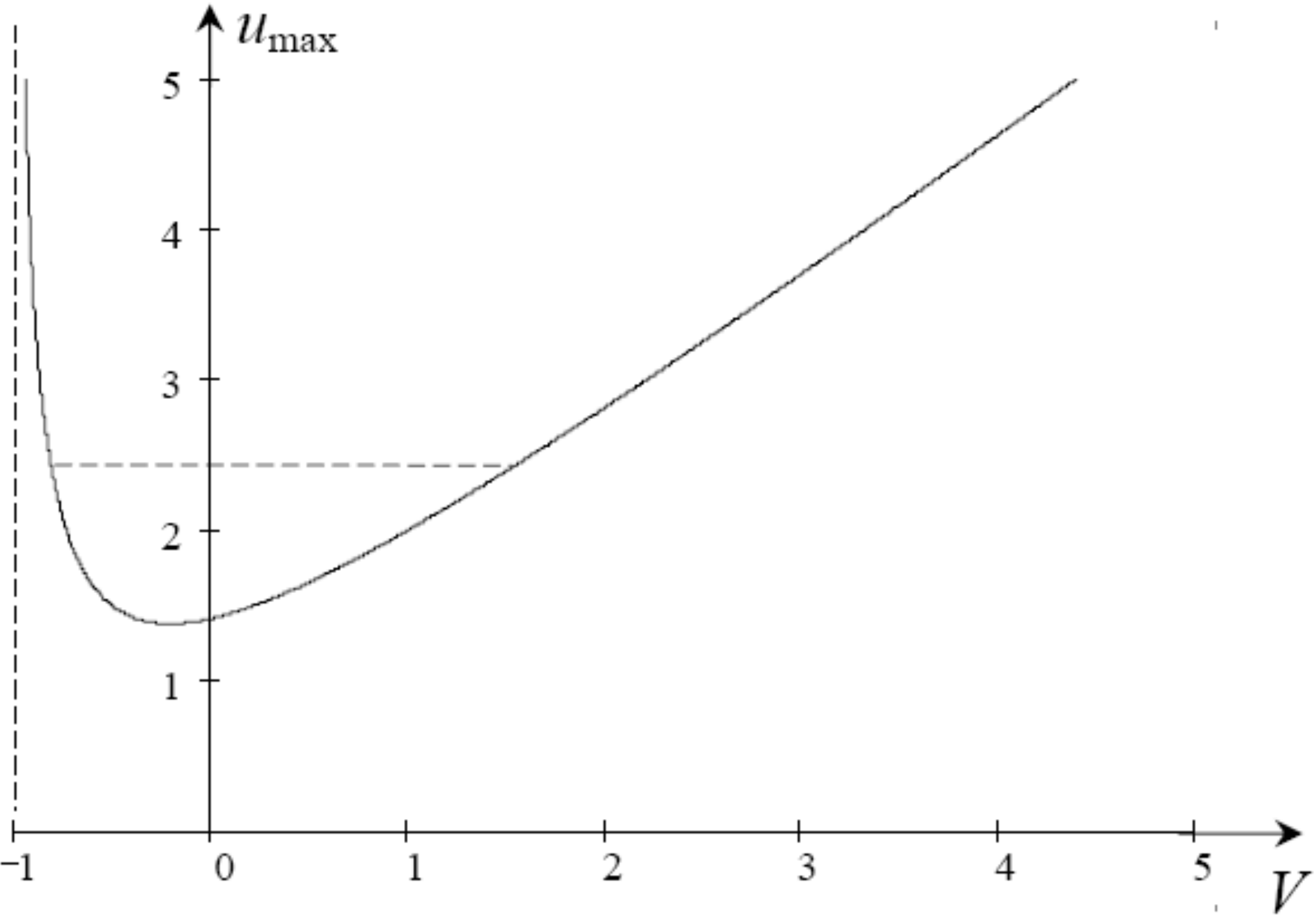}}
\caption{Dependence of the wave maximum on speed in original
variables, equation~\eqref{eq10}, as follows from solution~\eqref{eq9}.
Dashed vertical line corresponds to $V = -\beta/p$. The plot is
generated for $C_2 = p = \beta = 1$.}
\label{f06}%
\end{figure}

The maximum of the function $y(\xi)$, $y_{\max} = 1$, corresponds
in terms of $w$ to the wave maxi\-mum, $w_{\max} = (-C_2/E)^{1/2}$.
Using a relationship between $w$ and the original variable $u$
(see above), as well as def\/inition of the constant~$E$, one can
deduce the relationship between the wave maximum and its speed:
\begin{gather}
u_{\max} = V + \sqrt{\frac{-C_2}{E}} = V + \sqrt{\frac{2C_2}{pV +
\beta}}. \label{eq10}
\end{gather}
The plot of $u_{\max}(V)$ is presented in Fig.~\ref{f06} for $V >
-\beta/p$ in accordance with the chosen constant $C_2 = 1$ and $E
< 0$.

As follows from equation~\eqref{eq10}, wave maximum (minimum) cannot be
less than the certain value, $U_{\max}$ ($-U_{\min}$), which occurs
at some speed $V_1$, where
\begin{gather*}
U_{\max} = \frac{1}{p}\left[\left(\frac{C_2p^2}{2}\right)^{1/3} -
\beta\right] + 2\left(\frac{C_2}{2p}\right)^{1/3}, \qquad V_1 =
\frac{1}{p}\left[\left(\frac{C_2p^2}{2}\right)^{1/3} -
\beta\right]. 
\end{gather*}

For all possible values of wave maximum $u_{\max} > U_{\max}$, two
values of wave speed are pos\-sib\-le, i.e.\ two waves of the
very same ``amplitude'' can propagate with dif\/ferent speeds. This
is illustrated by horizontal dashed line in Fig.~\ref{f06} drawn
for $u_{\max} = 2.5$. The same is true for waves of negative
polarity.

{\bf 3c.} Consider now the case when both $C_1$ and $C_2$
are nonzero but $p + q$ is still zero. There are in general four
possible combinations of signs of the parameters $C_1$ and $C_2$:
\begin{gather*}
{\rm i)} \ C_1 > 0 , \ C_2 > 0;   \quad   {\rm ii)} \ C_1 < 0 , \ C_2 > 0; \quad
{\rm iii)} \ C_1 > 0 , \ C_2 < 0 ; \quad {\rm iv)} \ C_1 < 0 ,  \ C_2 < 0.
\end{gather*}

\begin{figure}[t]
\centerline{\includegraphics[width=150mm]{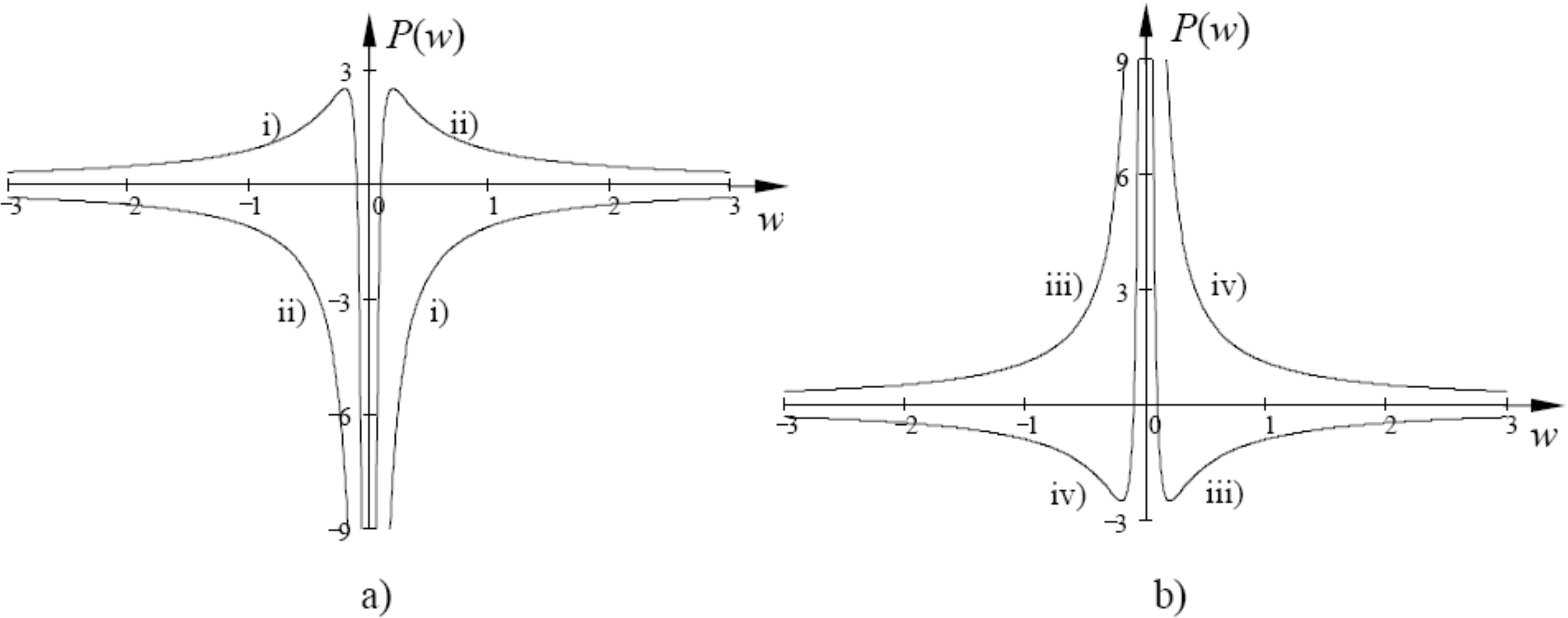}}
\caption{Potential relief for the four dif\/ferent cases, i)--iv),
of various signs of constants $C_1$ and $C_2$. The plot was
generated for $C_1 = \pm 1$, $C_2 = \pm 0.1$.}
\label{f07}%
\end{figure}

The shape of the potential function $P(w)$ and corresponding
solutions are dif\/ferent for all these cases. However, among them
there are only two qualitatively dif\/ferent and independent cases,
whereas the two others can be obtained from those two cases using
simple symmetry reasons. This statement is illustrated by
Fig.~\ref{f07}, where the potential relief is shown for all four
aforementioned cases i)--iv).

As one can see from Fig.~\ref{f07}, cases i) and ii), as well as
iii) and iv), are mirror symmetrical counterparts of each other
with respect to the vertical axis. This implies that solutions for
the cases i) and ii), and correspondingly, iii) and iv), are
related by the simple sign interchange operation, i.e.\
$w_{\text{i)}} = -w_{\text{ii)}}$, $w_{\text{iii)}} = -w_{\text{iv)}}$. Therefore, below only two
qualitatively dif\/ferent cases are considered in detail, namely the
cases~i) and~iii).

\begin{figure}[t]
\centerline{\includegraphics[height=85mm]{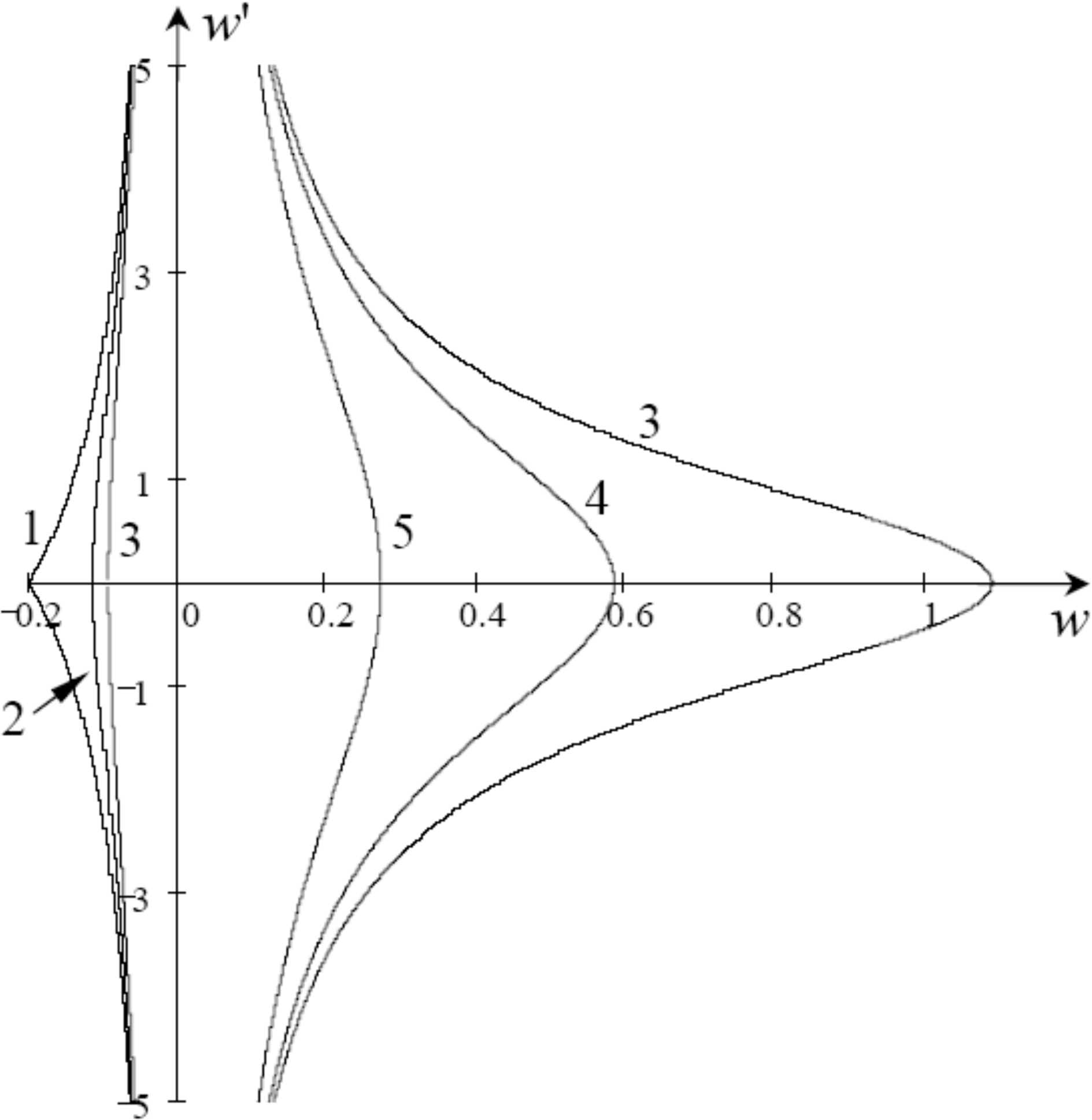}}
\caption{Phase portrait of equations~\eqref{eq4}, \eqref{eq5} for the
case i) (only those trajectories are shown which
correspond to particle motion within the potential well in
Fig.~\ref{f07}a). Line~1: $E = 2.5$; line~2: $E = 1$; lines~3: $E
= -1$; line 4: $E = -2$; line 4: $E = -5$.}
\label{f08}%
\end{figure}

Case i) is characterized by an inf\/inite potential well at the
origin, $w = 0$. This singularity in the potential function
corresponds to the existence of a singular straight line $w = 0$
on the phase plane (see Fig.~\ref{f08}). On both sides from this
singular line there are qualitatively similar trajectories which
correspond to bounded solutions having inf\/inite derivatives at the
edges. Quantitative dif\/ference between the ``left-hand side
solutions'' and ``right-hand side solutions'', apart of their
dif\/ferent polarity, is the former solutions (of negative polarity,
$w \le 0$) exist for $E \le P_{\max}$, whereas the latter ones (of
positive polarity, $w \ge 0$) exist for $E \le 0$. The potential
function has a maximum $P_{\max} = C_1^2/(4C_2)$ at $w =
-2C_2/C_1$. There are no bounded solutions for $E > P_{\max}$.

Consider f\/irst bounded solutions which correspond to trajectories
shown in the left half-plane, $w \le 0$, in Fig.~\ref{f08}. For a
positive value of the parameter $E$ in the range $0 \le E \le
P_{\max}$, the analytical solution can be presented in the form{\samepage
\begin{gather}
\xi(y) = \pm 2\sqrt{Q}\!\left[\!\!\sqrt{(y + 2Q)^2\! - 4Q(Q - 1)} -
2Q\ln{\frac{y + 2Q + \sqrt{(y + 2Q)^2\! - 4Q(Q - 1)}}{2\sqrt{Q(Q -
1)}}}\!\right]\!,\!\! \label{eq12}
\end{gather}
where $\xi = \chi\sqrt{2C_2}(C_1/C_2)^2$,  $y = w(C_1/C_2)$,
 $Q = C_1^2/(4C_2E)$.}

The range of variability on $\xi$ is:
\[
|\xi| \le 4Q\left\{1 -
\sqrt{Q}\ln{\left[\big(\sqrt{Q} + 1\big)/\sqrt{Q -
1}\right]}\right\},
\]
 whereas $y$ varies in the range
 \[
 -2\left[Q -
\sqrt{Q(Q - 1)}\right] \le y \le 0.
\]

The relationship between the wave minimum and its speed is:
\begin{gather}
u_{\min} = V + \frac{C_1}{pV + \beta} + \sqrt{\frac{2C_2}{pV +
\beta}\left[\frac{C_1^2}{2C_2(pV + \beta)} + 1\right]},
\label{eq13}
\end{gather}
where $pV + \beta < 0$ as $E > 0$.

If $E < 0$, then the solution is
\begin{gather}
\xi(y) = \pm 2\sqrt{-Q}\Bigg[\sqrt{4Q(Q - 1) - (y + 2Q)^2}\nonumber\\
\phantom{\xi(y) =}{} +
2Q\arctan{\left(\frac{y + 2Q}{\sqrt{4Q(Q - 1)  - (y +
2Q)^2}}\right) + \pi Q}\Bigg]. \label{eq14}
\end{gather}

The range of variability on $\xi$ is: $|\xi| \le -4Q\left[1 +
\sqrt{-Q}\left(\arctan{\sqrt{-Q} - \pi/2}\right)\right]$, whereas
$y$ varies in the range $-2\left[Q + \sqrt{Q(Q - 1)}\right] \le y
\le 0$. The relationship between the wave minimum and its speed is
also given by equation~\eqref{eq13}, but with $pV + \beta > 0$.

Two special cases of solution \eqref{eq12} can be mentioned. When
$Q = 1$ ($E = P_{\max}$), solution~\eqref{eq12} with the
appropriate choice of the integration constant reduces to
\begin{gather}
\xi(y) = \pm 4\left[\frac{y}{2} - \ln{\left(1 +
\frac{y}{2}\right)}\right]. \label{eq15}
\end{gather}

This solution is unbounded on $\xi$, i.e.\ it is def\/ined in
the range: $|\xi| \le \infty$. However, the solution is bounded on
$y$: $-2 \le y \le 0$. The relationship between the wave minimum
and its speed is simple as both of them are constant values in
this special case:
\begin{gather}
V = -\frac{1}{p}\left(\beta + \frac{C_1^2}{2C_2}\right), \qquad
u_{\min} = V - 2\frac{C_2}{C_1} = -\frac 1p\left(\beta +
\frac{C_1^2}{2C_2} + 2p\frac{C_2}{C_1}\right). \label{eq16}
\end{gather}

Another special case corresponds to $Q = \infty$ ($E = 0$); in
this case equation~\eqref{eq12} after appropriate choice of integration
constant reduces to:
\begin{gather}
\xi(y) = \pm\frac 23\sqrt{y +1}(y - 2). \label{eq17}
\end{gather}

The range of variability on $\xi$ is: $|\xi| \le 4/3$, whereas $y$
varies in the range: $-1 \le y \le 0$. The relationship between
the wave minimum and its speed is also very simple as both of them
are again constants but dif\/ferent from those given by
equation~\eqref{eq16}; in this case they are:
\begin{gather*}
V = -\frac{\beta}{p}, \qquad u_{\min} = V - \frac{C_2}{C_1} =
-\left(\frac{\beta}{p} + \frac{C_2}{C_1}\right). 
\end{gather*}

Bounded solutions corresponding to the trajectories shown in the
right half-plane in Fig.~\ref{f08} with $w \ge 0$, exist only for
negative $E$; they are given by equation~\eqref{eq14}, but with the
dif\/ferent range of variability of $y$: $0 \le y \le 2\left[-Q +
\sqrt{Q(Q - 1)}\right]$. The relationship between the wave maximum
and its speed is given again by equation~\eqref{eq13} where $u_{\max}$
should be substituted instead of $u_{\min}$ and $pV + \beta > 0$ as
$E < 0$ for these solutions.

\begin{figure}[t]
\centerline{\includegraphics[height=72mm]{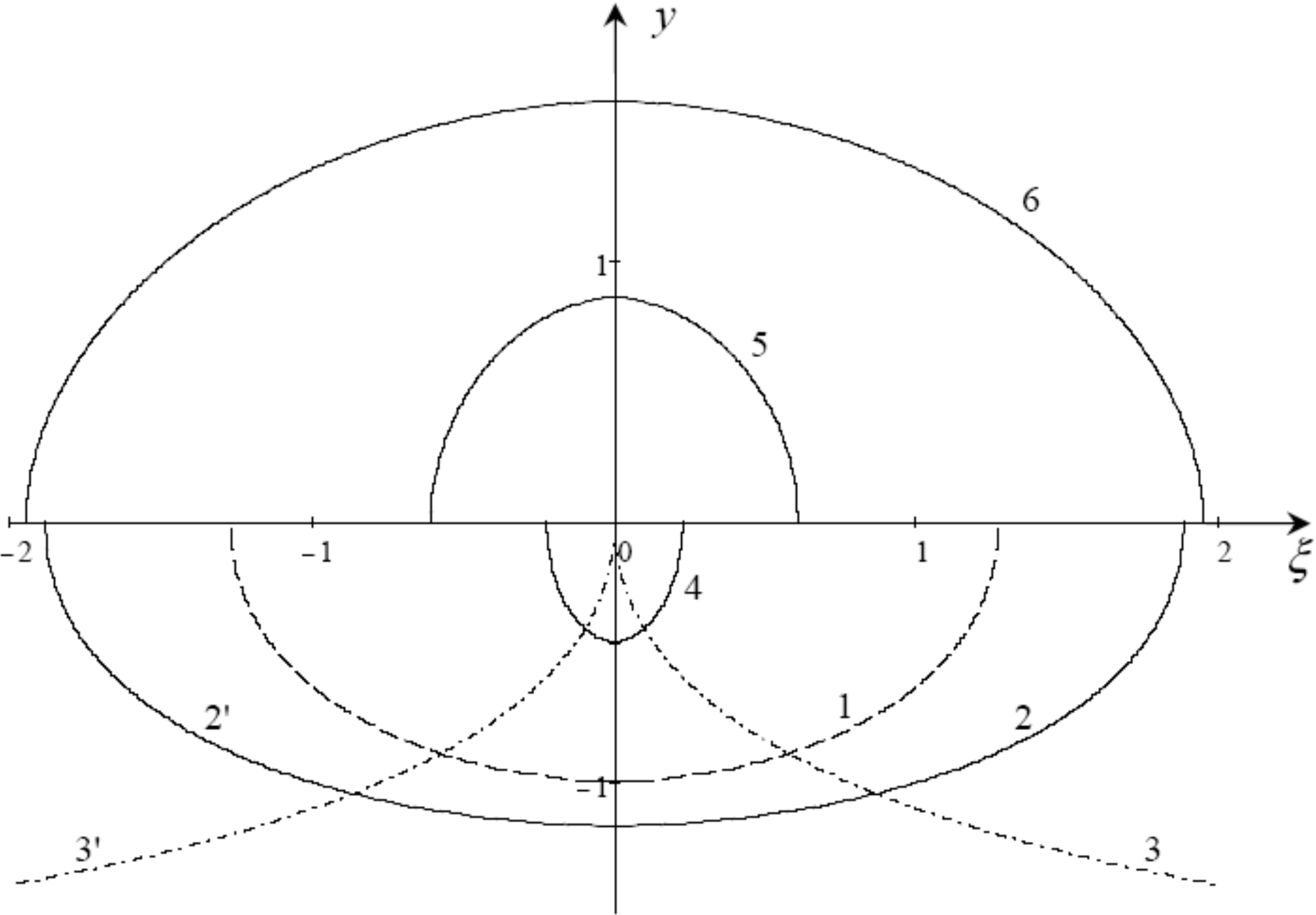}}

\caption{Various solutions described by equations~\eqref{eq12},
\eqref{eq14}, \eqref{eq15} and \eqref{eq17}. Compactons of
negative polarity: line 1: $Q = \infty$; lines 2 and $2'$: $Q =
2$; lines 3 and $3'$: $Q = 1$:  line 4: $Q = -0.1$. Compactons of
positive polarity: line 5: $Q = -0.1$; line 6: $Q = -0.25$.}
\label{f09}%
\end{figure}

Solutions \eqref{eq12}, \eqref{eq14}, \eqref{eq15} and
\eqref{eq17} are shown in Fig.~\ref{f09}. All these solutions are
of the compacton type; they consist of two independent branches
which can be matched dif\/ferently or unmatched at all. Lines 2 and
$2'$ represent an example when two branches are matched so that
they form a semi-oval; lines~3 and~$3'$ represent another example
when two branches are matched so that they form an inverted
``seagull''. On the basis of these ``elementary'' solutions,
various complex compound solutions can be constructed including
periodic or chaotic stationary waves.

The dashed line 1 in the f\/igure corresponds to $E = 0$ ($Q = 8$).
Another branch of the solution with the same value of $E = 0$
represents a solution of positive polarity which is unbounded both
on $\xi$ and $y$. For positive values of $E$, solutions of
negative polarity become wider and of greater ``amplitude'' (see
line~2). When $E$ further increases and approaches $P_{\max}$, the
solution becomes inf\/initely wide, but its minimum goes to $-2$. In
the limiting case $E = P_{\max}$ ($Q = 1$) two independent branches
of the solution can be matched dif\/ferently as shown by
dashed-dotted lines 3 and $3'$ in Fig.~\ref{f09}. The solution
vanishes in this case when $\xi = 0$ and goes to $-2$ when $\xi
\to \pm\infty$; this situation is described by equation~\eqref{eq15}.

For the negative $E$ there are two families of solutions: negative
one, corresponding to the left-hand side trajectories in
Fig.~\ref{f08}, and positive one, corresponding to the right-hand
side trajectories. When $E$, being negative, increases in absolute
value ($Q$ varies from $-\infty$ to $0_-$), solutions depart from
the line~1 in Fig.~\ref{f09} and gradually squeeze to the origin
(see line~4 for instance). For the same values of negative $E$,
positive solutions originated at inf\/inity also gradually shrink
and collapse in the origin (lines 6 and 5 demonstrate this
tendency).

\begin{figure}[t]
\centerline{\includegraphics[height=75mm]{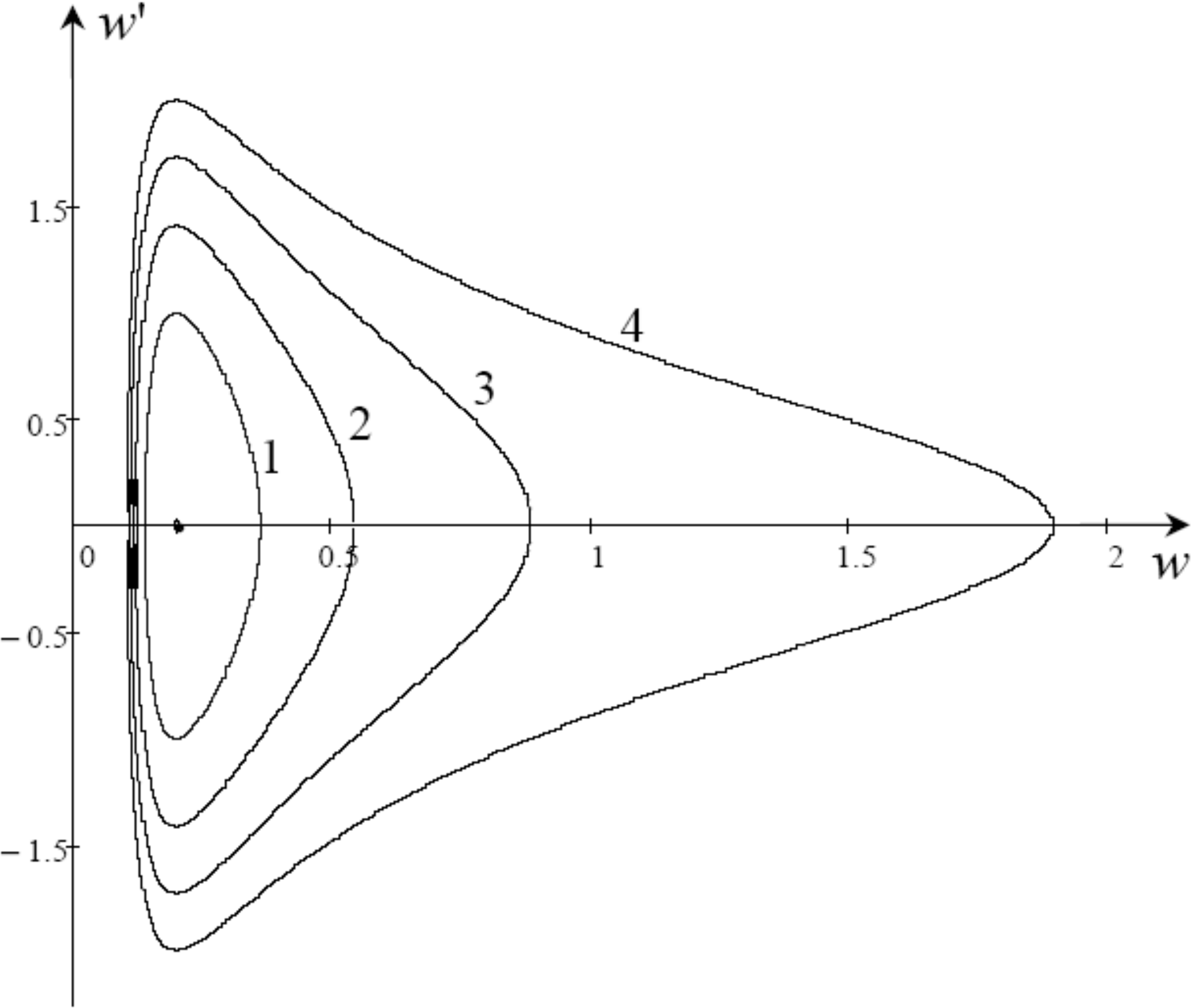}}
\caption{Phase portrait of equations~\eqref{eq4}, \eqref{eq5} for the
case {\bf 3c} iii) (only those trajectories are shown
which correspond to particle motion within the potential well in
Fig.~\ref{f07}b). The dot at the center of closed lines indicates
an equilibrium point corresponding to the potential minimum ($E =
-2.5$ for the chosen set of parameters: $C_1 = 1$, $C_2 = -0.1$);
line 1: $E = -2$; line 2: $E = -1.5$; line 3: $E = -1$; line 4: $E
= -0.5$.}
\label{f10}%
\end{figure}

Consider now the case iii) shown in Fig.~\ref{f07}b. The potential
function in this case has only one well of a f\/inite depth so that
$P_{\min} = C_1^2/(4C_2)$ at $w = -2C_2/C_1$, where $C_2$ is
negative now. There are no bounded solutions for negative w; they
exist however for positive $w$ and $E$ varying in the range
$P_{\min} \le E < 0$. The f\/inite value of the potential minimum
corresponds to the equilibrium point of the centre type in the
phase plane. There is also a family of closed trajectories for the
above indicated range of $E$ variation (see Fig.~\ref{f10}); these
trajectories correspond to periodic solutions.

\begin{figure}[t]
\centerline{\includegraphics[height=60mm]{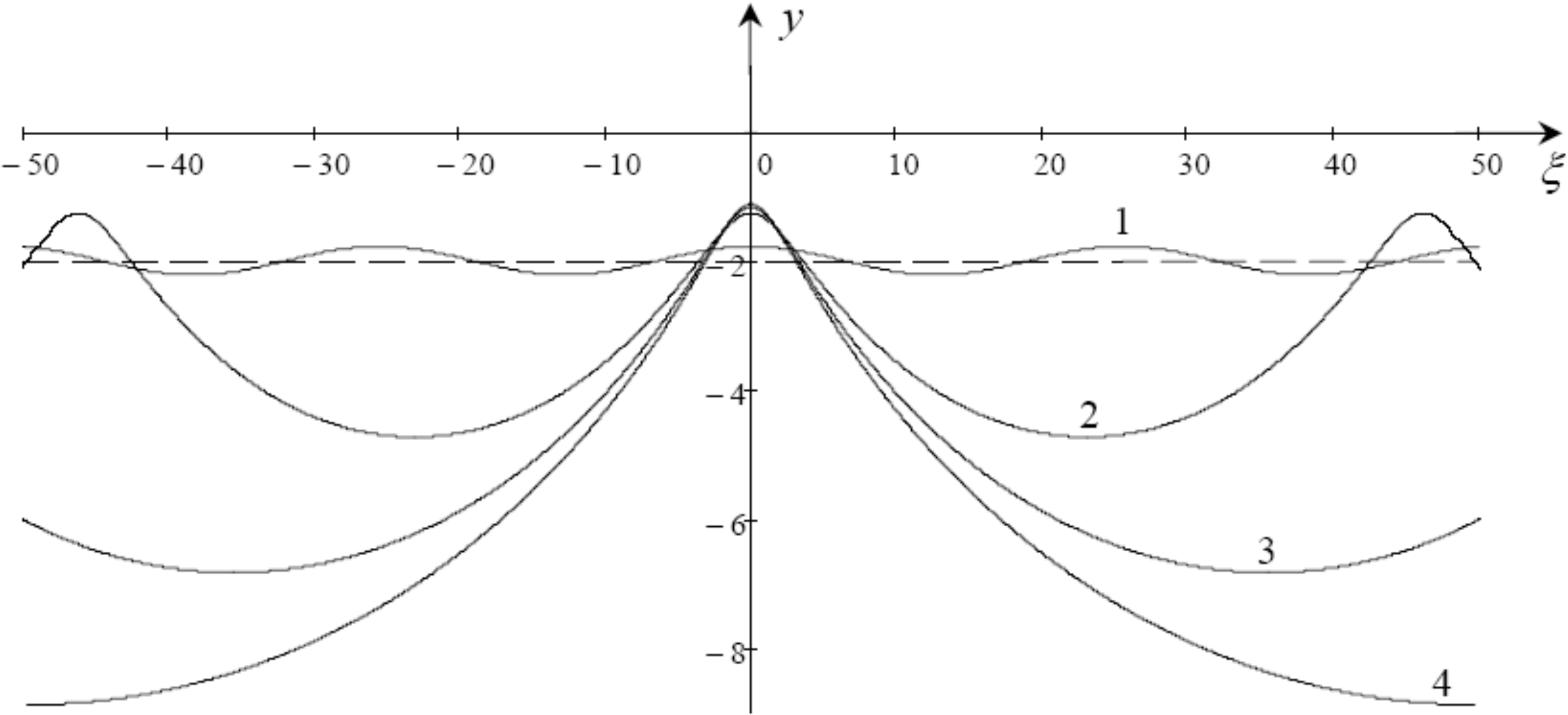}}

\caption{Various solutions described by equation~\eqref{eq19}. Line~1
(quasi-sinusoidal wave): $Q = 1.01$; line~2: $Q = 1.5$; line~3: $Q
= 2$; line~4: $Q = 2.5$. Dashed lines shows the equilibrium state
$y = -2$.}
\label{f11}%
\end{figure}

As usual, closed trajectories around the center ($E \ge P_{\min}$)
correspond to quasi-sinusoidal solutions. Whereas other closed
trajectories ($E > P_{\min}$) correspond to non-sinusoidal periodic
waves with smooth crests and sharp narrow troughs. The larger is
the value of $E$, the longer is the wave period. The period tends
to inf\/inity when $E \to 0_-$. The analytical form of this family
of solutions is described by the following equation:
\begin{gather}
\xi(y) = \pm 2\sqrt{Q}\Bigg[\sqrt{4Q(Q - 1) - (y + 2Q)^2}\nonumber\\
\phantom{\xi(y) =}{} +
2Q\arctan{\left(\frac{y + 2Q}{\sqrt{4Q(Q - 1)  - (y +
2Q)^2}}\right) - \pi Q}\Bigg], \label{eq19}
\end{gather}
where $\xi = \chi\sqrt{-2C_2}(C_1/C_2)^2$,  $y = w(C_1/C_2)$,
  $Q = C_1^2/(4C_2E)$. Solution \eqref{eq19} is shown in
Fig.~\ref{f11} for dif\/ferent values of $Q$ (note that the solution
is negative in terms of $y$ because $C_2 < 0$). As follows from
equation~\eqref{eq19}, $y$ varies in the range:
\[
-2\left[Q + \sqrt{Q(Q
- 1)}\right] \le y \le -2\left[Q - \sqrt{Q(Q - 1)}\right],
\]
whereas the dependence of wave period $\Lambda$ on $Q$ is:
$\Lambda(Q) = 8\pi Q\sqrt{Q}$. The wave period varies from $8\pi$
to inf\/inity when $Q$ increases from unity to inf\/inity.

\begin{figure}[t]
\centerline{\includegraphics[height=65mm]{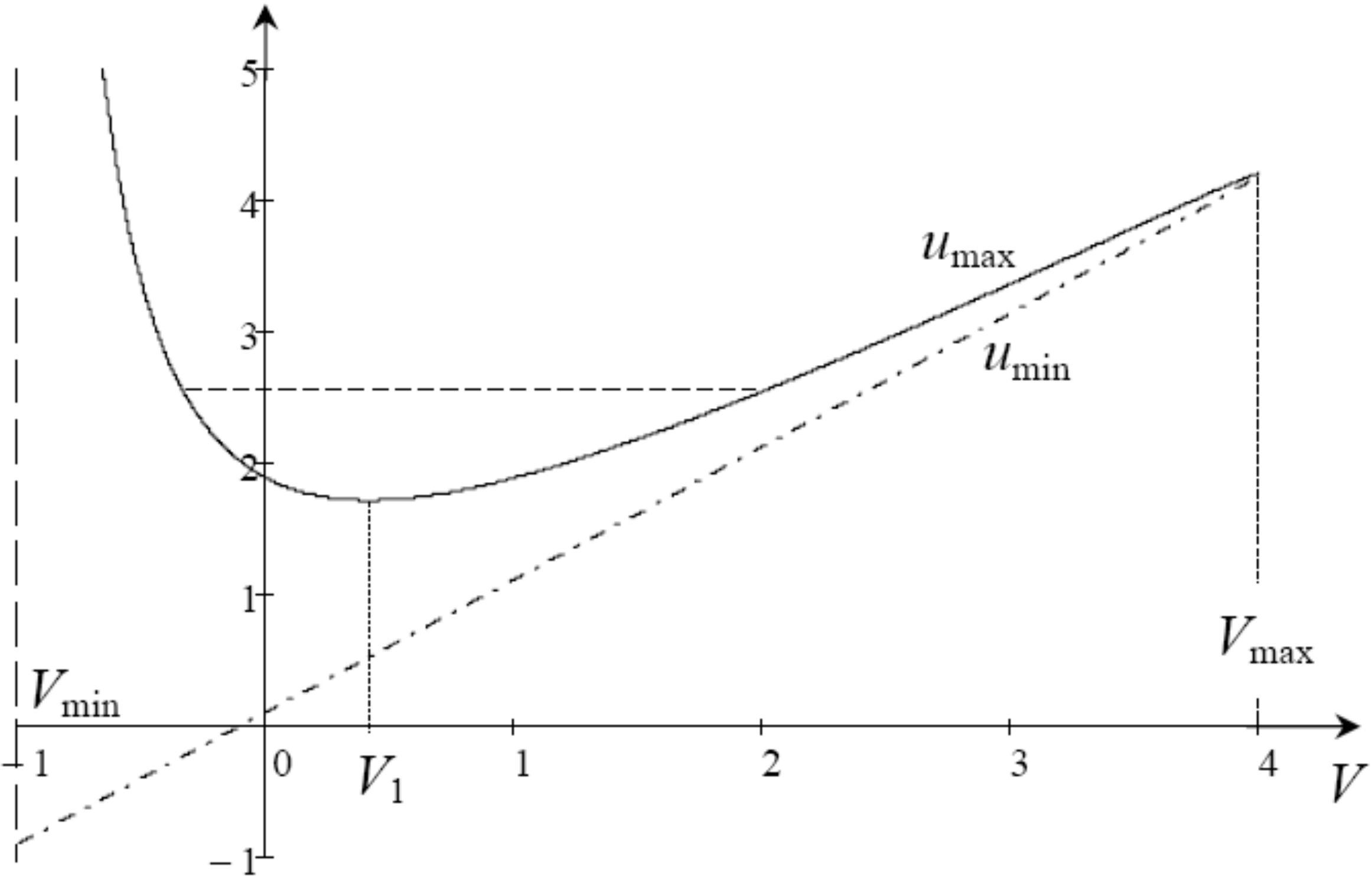}}

\caption{Dependences of wave maximum (solid line) and minimum
(dashed-dotted line) on speed in the original variables,
equation~\eqref{eq20}, as follows from the solution \eqref{eq19}.
Dashed vertical line corresponds to $V = -\beta/p$. The plot is
generated for $C_1 = 1$, $C_2 = -0.1$ and $p = \beta = 1$.}
\label{f12}%
\end{figure}

From the extreme values of $y$ (see above indicated range of its
variability) one can deduce the dependences of wave maximum and
minimum on speed in the original variables. The corresponding
formulae are:
\begin{gather}
u_{\max, \min}(V) = V + \frac{C_1}{pV + \beta} \mp
2C_2\sqrt{\frac{1}{2C_2(pV + \beta)}\left[\frac{C_1^2}{2C_2(pV +
\beta)} + 1\right]}, \label{eq20}
\end{gather}
where the upper sign in front of the root corresponds to the wave
maximum and lower sign -- to the wave minimum. These dependences
are plotted in Fig.~\ref{f12} for $V > -\beta/p$ in accordance
with the chosen values of constants $C_1 = 1$, $C_2 = -0.1$ and $E
< 0$. The asymptote $V = -\beta/p$ is shown in the f\/igure by the
vertical dashed line. As follows from equation~\eqref{eq17}, wave
maximum cannot be less than the certain value, $U_{\max}$, which
occurs at some speed $V_1$ shown in Fig.~\ref{f12}.

For all possible values of wave maximum $u_{\max} > U_{\max}$, two
values of wave speed are possible, i.e.\ two periodic waves
of the same maximum (but not minimum!) can propagate with
dif\/ferent speeds. This is illustrated by the horizontal dashed
line shown in Fig.~\ref{f12} and drawn for \mbox{$u_{\max} = 2.5$}. In
original variables quasi-sinusoidal waves exist when the speed is
close to its limiting value $V_{\max} = -\frac
1p\left(\frac{C_1^2}{2C_2} + \beta\right)$; there are no waves
with greater speed. When $V \le V_{\max}$, the wave minimum and
maximum are close to each other. Then, when the speed decreases,
the gap between wave maximum and minimum gradually increases and
goes to inf\/inity when the speed approaches its minimum value
$V_{\min} = -\beta/p$.

\section[General case: $p + q \ne 0$]{General case: $\boldsymbol{p + q \ne 0}$}
\label{sec4} %

Consider now a more general case when the coef\/f\/icients in
equation~\eqref{eq1} are such that $p + q \ne 0$. The basic equation
\eqref{eq5} can be presented in the new variables $\eta = (p +
q)\chi/6$ and $v = (p + q)w/6$ with the same constant of
integration $E = -(pV + \beta)/2$, but with new ef\/fective
potential function
\begin{gather}
P(v) = v - \frac{C_1}{v} - \frac{C_2}{v^2}. \label{eq22}
\end{gather}

The potential function is monotonic when $C_1 = C_2 = 0$, and
there are no bounded solutions in this case. Bounded solutions may
exist if at least one of these constants is nonzero. Below we
present possible forms of the potential function and corresponding
phase portraits of bounded solutions for various relationships
between constants $C_1$ and $C_2$. Qualitatively all these cases
are similar to those which have been described already in the
previous section, therefore we omit the detailed analysis and do
not present analytical solutions as they can be obtained
straightforwardly and expressed in terms of elliptic functions.

\begin{figure}[t]
\centerline{\includegraphics[height=60mm]{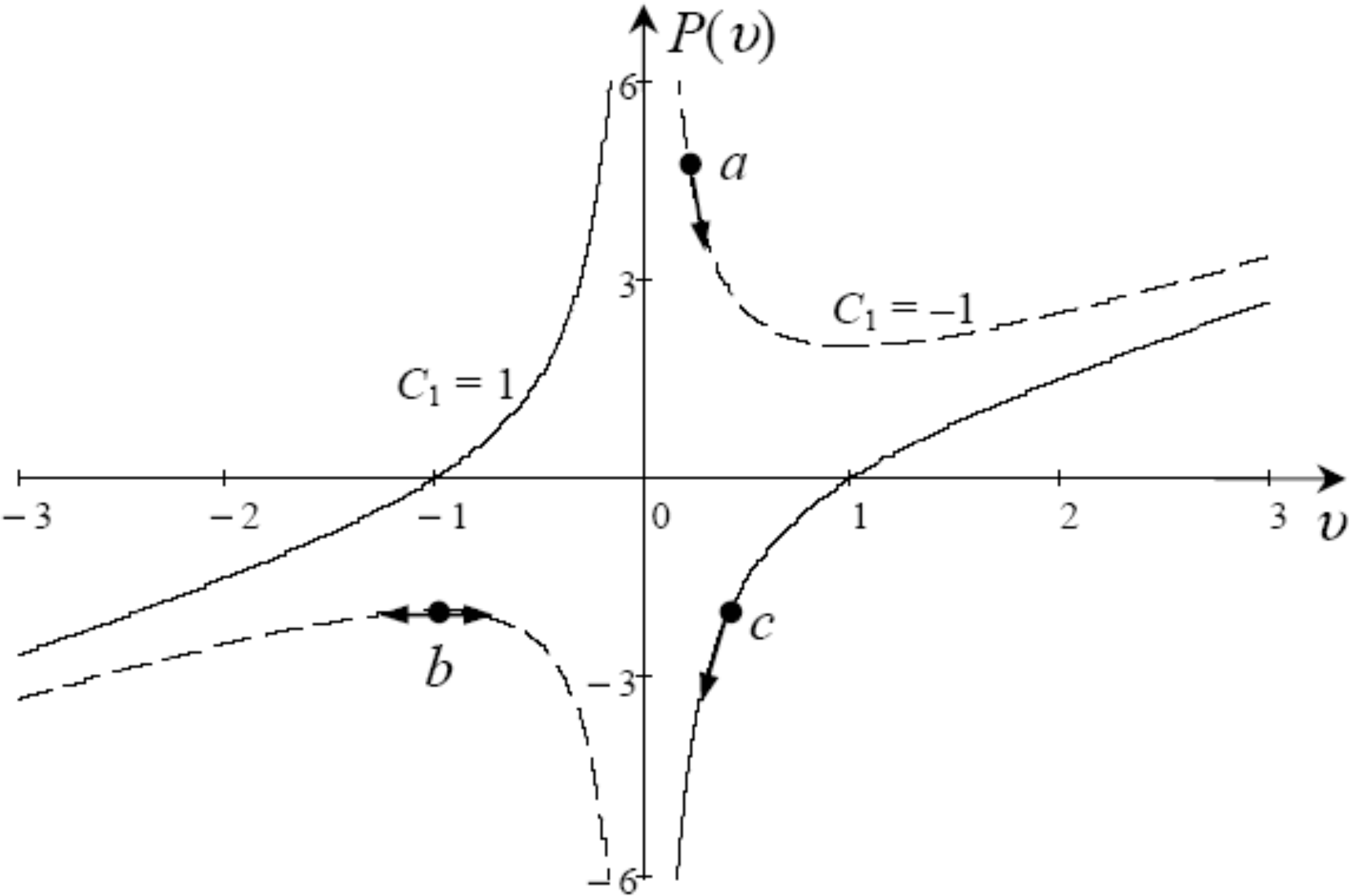}}

\caption{Potential function for the case $p + q \ne 0$, $C_2 = 0$
and two values of $C_1$: $C_1 = 1$ (solid line), and $C_1 = -1$
(dashed line). Dots $a$, $b$ and $c$ illustrate possible motion of
a point particle in the potential f\/ield.}
\label{f13}%
\end{figure}

{\bf 4a}. If $C_2 = 0$, the potential function represents
a set of antisymmetric hyperbolas located either in the f\/irst and
third quadrants when $C_1 = -1$, or in the second and fourth
quadrants when $C_1 = 1$; this is shown in Fig.~\ref{f13}.

\begin{figure}[t]
\centerline{\includegraphics[width=150mm]{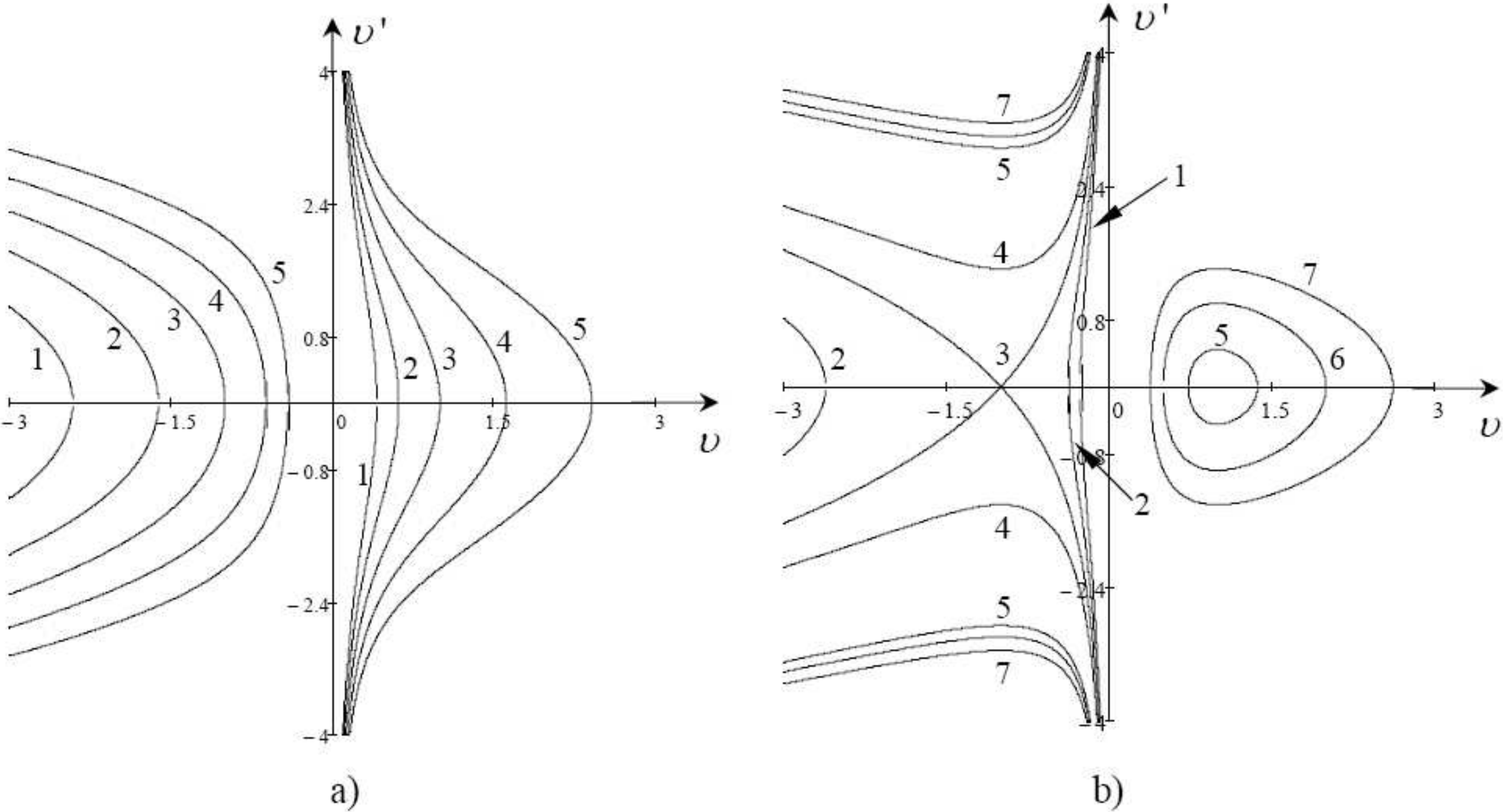}}

\caption{a) Phase plane corresponding to the potential function
with $C_1 = 1$ and various values of~$E$. Line~1: $E = -2$; lines~2: $E = -1$; lines~3: $E = 0$; lines 4: $E = 1$; lines 5: $E = 2$.
All trajectories in the left half-plane correspond to unbounded
solutions. b) Phase plane corresponding to the potential function
with $C_1 = -1$ and various values of $E$. Line 1: $E = -4$; lines
2: $E = -3$; lines~3: $E = -2$; lines~4: $E = -1$; lines 5: $E =
2.1$; lines 6: $E = 2.5$; lines 7: $E = 3$.}
\label{f14}%
\end{figure}

For the case of $C_1 = 1$ only bounded solutions of a compacton
type are possible for positive $v$. Such solutions correspond to
the motion of particle $c$ shown in the f\/igure down to the potential
well. This family of pulse-type solutions exist both for negative
and positive $E$; all of them are bounded from the top with the
maximum values depending on $E$, have zero minimum values and
inf\/inite derivatives when $v = 0$. Corresponding phase plane is
presented in Fig.~\ref{f14}a.

For the case of $C_1 = -1$ there are two possibilities: i) there
is a family of compacton-type solutions with $v \le 0$; they
correspond to the motion of the particle $b$ down to the potential
well (particle motion to the left from the top of the ``hill''
corresponds to unbounded solutions). Possible values of particle
energy $E$ vary for such motions from minus inf\/inity to $P_{\max} =
-2\sqrt{-C_1}$, where $P_{\max}$ is the local maximum of the lower
branch of the potential function (see Fig.~\ref{f13}). The phase
portrait of such motions is shown in the left half of the phase
plane in Fig.~\ref{f14}b.

ii) Another possibility appears for the particle motion within the
potential well shown in the f\/irst quadrant of Fig.~\ref{f13} (see
the particle $a$). Within this well all phase trajectories are
closed and corresponding solutions are bounded and periodical;
they can be expressed in terms of elliptic functions. The phase
portrait of such motions is shown in the right half of the phase
plane in Fig.~\ref{f14}b.

\begin{figure}[t]
\centerline{\includegraphics[height=60mm]{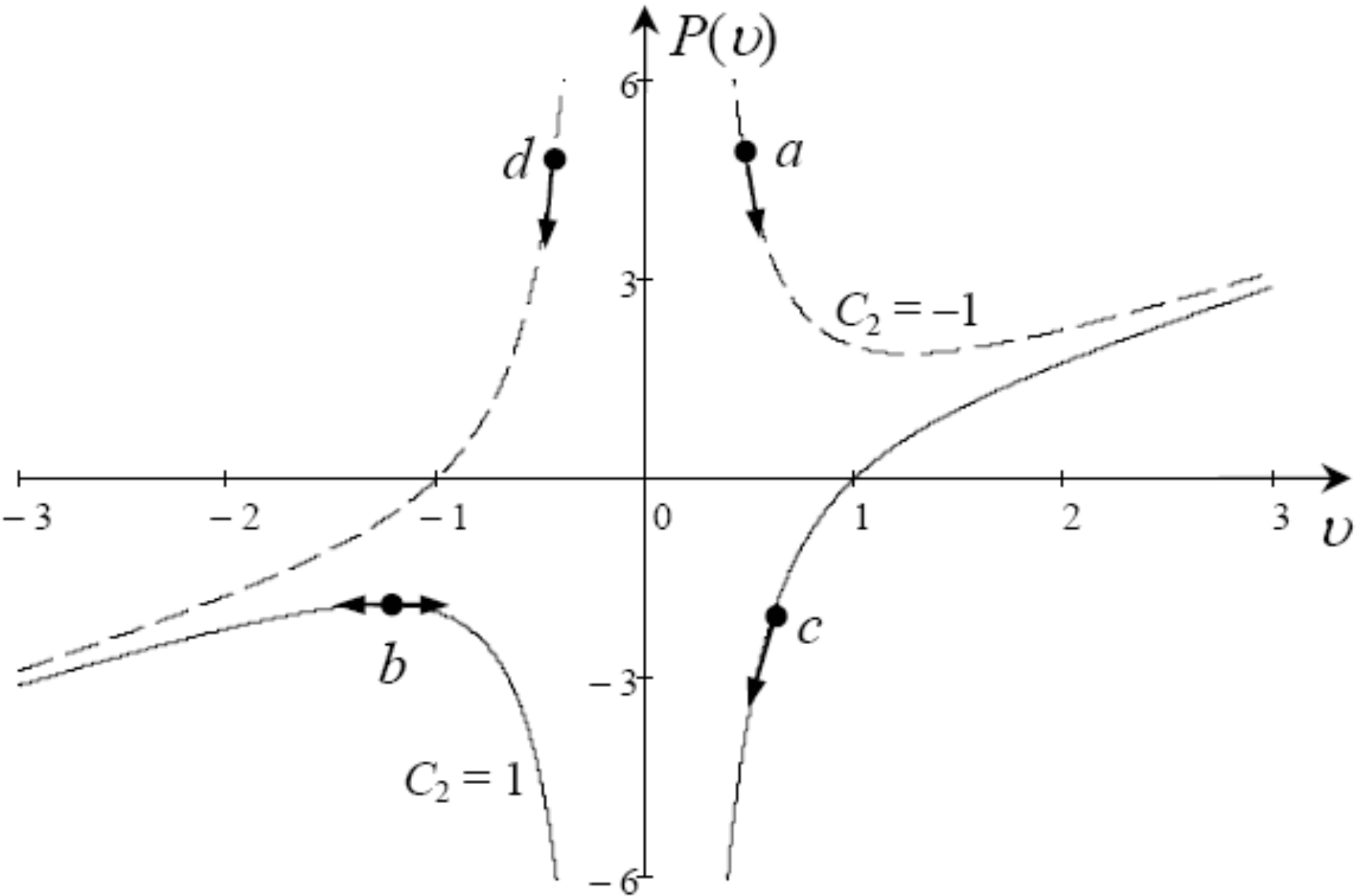}}

\caption{Potential function for the case $p + q \ne 0$, $C_1 = 0$
and two values of $C_2$: $C_2 = 1$ (solid lines), and $C_2 = -1$
(dashed lines). Dots $a$, $b$, $c$ and $d$ illustrate possible
motion of a point particle in the potential f\/ield.}
\label{f15}%
\end{figure}

{\bf 4b}. If $C_1 = 0$, but $C_2 \ne 0$, the potential
function also represents a set of antisymmetric hyperbolas located
either in the third quadrant and right half-plane in
Fig.~\ref{f15} when $C_2 = 1$, or in the f\/irst quadrant and left
half-plane of that f\/igure when $C_2 = -1$.

\begin{figure}[t]
\centerline{\includegraphics[width=150mm]{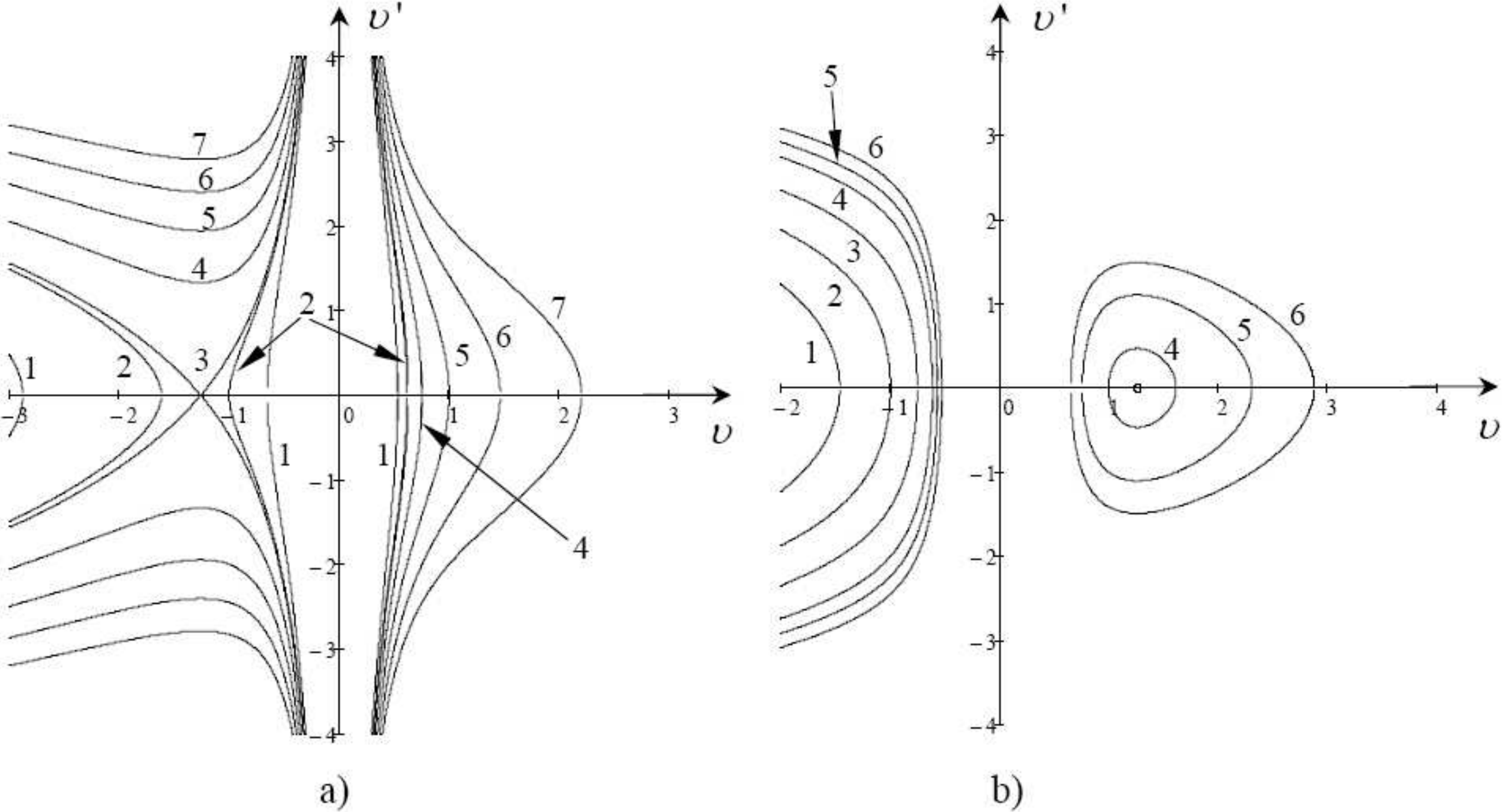}}

\caption{Phase plane corresponding to the potential function
\eqref{eq22} with $C_1 = 0$. a) $C_2 = 1$ and various values of
$E$. Lines 1: $E = -3$; lines 2: $E = -2$; lines 3: $E = -1.89$;
lines 4: $E = -1$; lines 5: $E = 0$; lines~6: $E = 1$; lines 7: $E
= 2$. b) $C_2 = -1$ and various values of~$E$. Line 1: $E = -1$;
line 2: $E = 0$; line 3: $E = 1$; lines 4: $E = 2$; lines 5: $E =
2.5$; lines 6: $E = 3$. All trajectories in the left half-plane
correspond to unbounded solutions.}
\label{f16}%
\end{figure}

For the case of $C_2 = 1$ there are two possibilities: i) there is
a family of compacton-type solutions with $v \le 0$; they
correspond to the motion of the particle $b$ down to the potential
well (particle motion to the left from the top of the ``hill''
corresponds to unbounded solutions). Possible values of particle
energy $E$ vary for such motions from minus inf\/inity to $P_{\max} =
3(-C_2/4)^{1/3}$, where $P_{\max}$ is the local maximum of the left
branch of the potential function (see Fig.~\ref{f15}). The phase
portrait of such motions is shown in the left half-plane in
Fig.~\ref{f16}a.

ii) Another family of compacton-type solutions exist with $v \ge
0$; they correspond to the motion of the particle $c$ down to the
potential well. Possible values of particle energy $E$ for such
motions vary from minus to plus inf\/inity. Corresponding phase
plane is presented in the right half-plane in Fig.~\ref{f16}a.

For the case of $C_2 = -1$ bounded solutions are smooth periodic
waves which correspond to the particle oscillations in the
potential well shown in the f\/irst quadrant in Fig.~\ref{f15}.
Energy is positive for such motion and varies from $P_{\min} =
3(-C_2/4)^{1/3}$, where $P_{\min}$ is the local minimum of the
right branch of the potential function (see Fig.~\ref{f15}) to
inf\/inity. Analytical solution for such waves can be also expressed
in terms of cumbersome elliptic functions. Corresponding phase
plane is presented in Fig.~\ref{f16}b.

{\bf 4c}. Consider now the case when both $C_1 \ne 0$ and
$C_2 \ne 0$. The shape of the potential function is more complex
in this case in general and depends on the relationship between
the constants $C_1$ and $C_2$. The number and values of the
potential extrema are determined by the number of real roots of
the equation $P'(v) = 0$, where prime denotes the derivative on
$v$. This condition yields (see equation~\eqref{eq22}):
\begin{gather*}
v^3 + C_1v + 2C_2 = 0. 
\end{gather*}

For real constants $C_1$ and $C_2$ this equation always has at
least one real root. The real root is single when $C_1 \ge
C_1^{\rm cr} \equiv -3C_2^{2/3}$; its value is given by the expression
\begin{gather*}
v = \left[\sqrt{(C_1/3)^3 + C_2^2} - C_2\right]^{1/3} -
(C_1/3)\left[\sqrt{(C_1/3)^3 + C_2^2} - C_2\right]^{-1/3}.
\end{gather*}

\begin{figure}[t]
\centerline{\includegraphics[width=145mm]{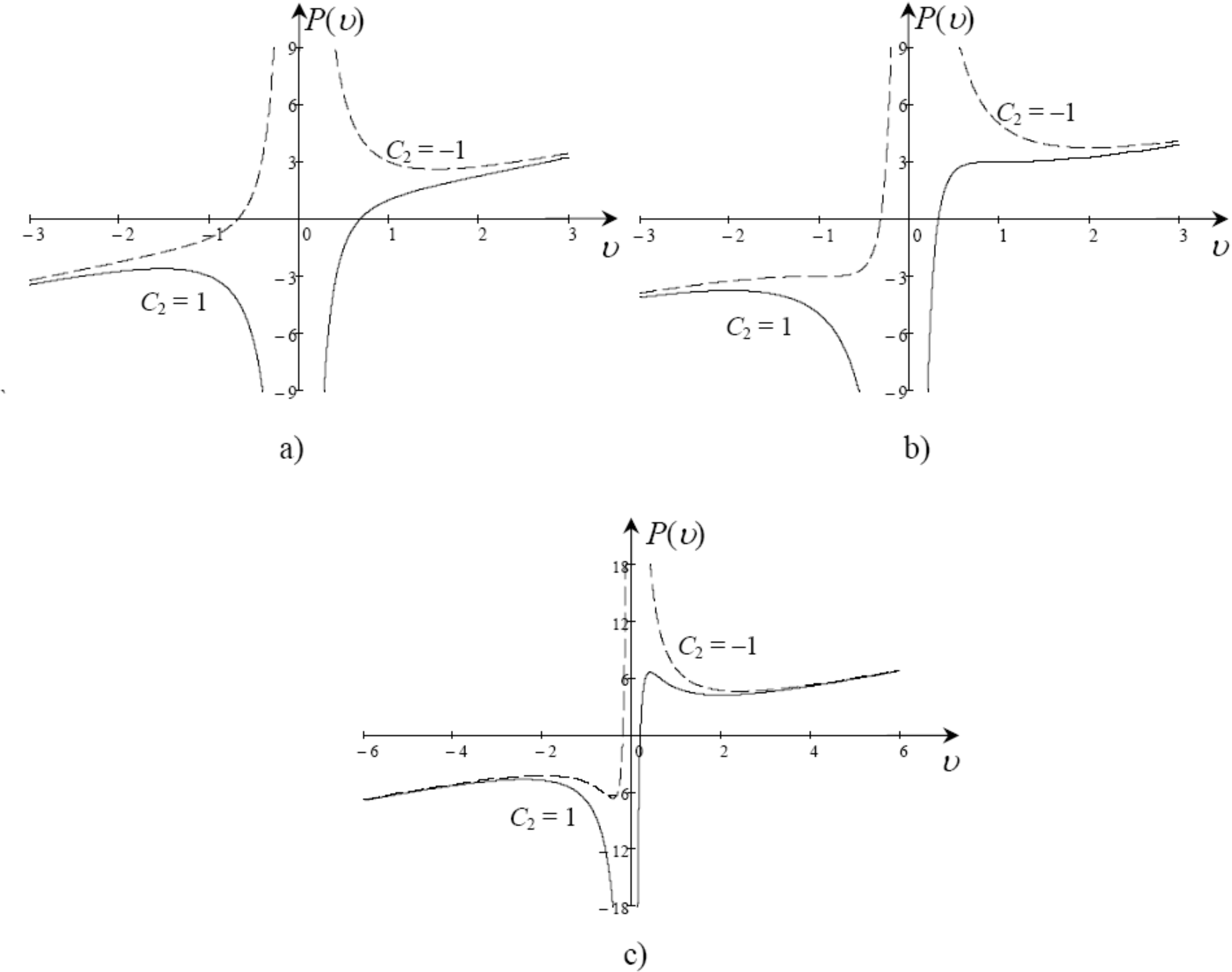}}

\caption{Potential function for the case $p + q \ne 0$. a)
Supercritical case: $C_1 = -1$ and two values of $C_2$: $C_2 = 1$
(solid lines), and $C_2 = -1$ (dashed lines); b) marginal case:
$C_1 = -3$ and the same two values of $C_2$; c) subcritical case:
$C_1 = -5$ and the same two values of $C_2$ (in the last case the
horizontal and vertical scales are doubled).}
\label{f17}%
\end{figure}

For the case $C_1 > C_1^{\rm cr}$, possible qualitative conf\/igurations
of the potential function are shown in Fig.~\ref{f17}a for the
particular choices of constants: $C_1 = -1$ and $C_2 = \pm 1$.
There is only one local minimum at the right branch of the
potential function for $C_2 = -1$ and a local maximum at the left
branch of the potential function for $C_2 = 1$. Almost the same
conf\/iguration of the potential function occurs for the marginal
case $C_1 = C_1^{\rm cr}$, as shown in Fig.~\ref{f17}b, however one
more local extremum appears -- on the left branch when $C_2 = -1$
and on the right branch when $C_2 = 1$. In the case $C_1 <
C_1^{\rm cr}$ the potential function is shown in Fig.~\ref{f17}c;
there are three local extrema of the potential function for any
value of $C_2 = \pm 1$.

\begin{figure}[t]
\centerline{\includegraphics[width=145mm]{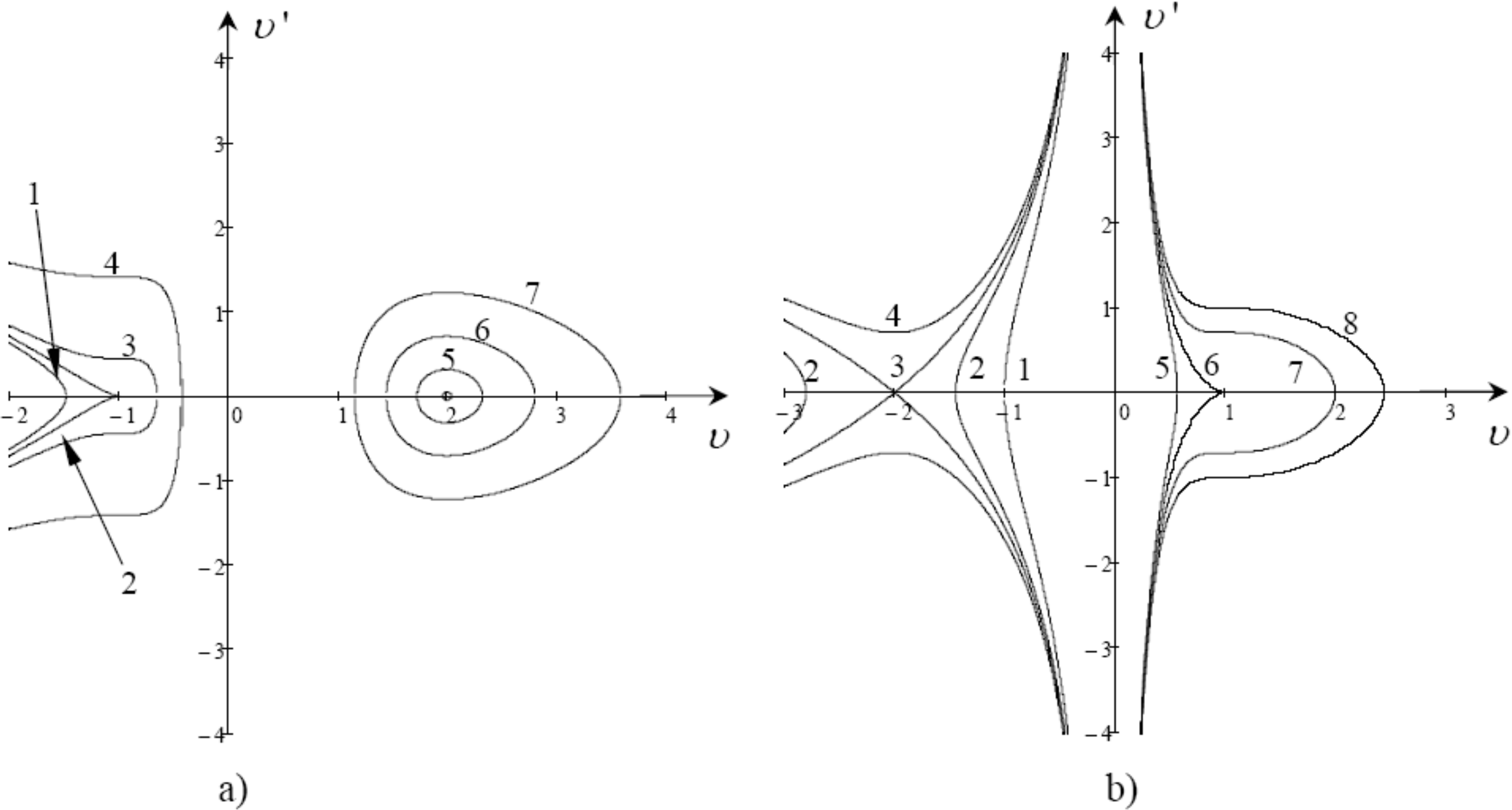}}

\caption{Phase plane corresponding to the marginal case, $C_1 =
C_1^{\rm cr}$. a) $C_1 = -3$, $C_2 = -1$. Line 1: $E = -3.05$; line 2:
$E = -3$; line 3: $E = -2.9$; line 4: $E = -2$; line 5: $E = 3.8$;
line 6: $E = 4$; line 7: $E = 4.5$. All trajectories in the left
half-plane correspond to unbounded solutions. b) $C_1 = -3$, $C_2
= 1$. Line 1: $E = -5$; lines 2: $E = -4$; lines 3: $E = -3.75$;
lines 4: $E = -3.5$; line 5: $E = 2.75$; line 6: $E = 3$; line 7:
$E = 3.25$; line 8: $E = 3.5$.}
\label{f18}%
\end{figure}

The potential conf\/iguration in the supercritical case $C_1 >
C_1^{\rm cr}$ qualitatively is similar to the case shown in
Fig.~\ref{f15}, therefore the corresponding phase portraits are
similar to those shown in Fig.~\ref{f16}. In the marginal case,
$C_1 = C_1^{\rm cr}$, the potential conf\/iguration is also similar to
those two cases mentioned above, however there are some
peculiarities in the phase planes ref\/lecting the appearance of
embryos of new equilibrium points. Corresponding phase portraits
are shown in Fig.~\ref{f18}. The embryos appear in the vicinity of
$E = -3$ in Fig.~\ref{f18}a and in the vicinity of $E = 3$ in
Fig.~\ref{f18}b.

In the subcritical case $C_1 < C_1^{\rm cr}$ the situation is
dif\/ferent from the previous ones and should be considered
separately. In the case of $C_2 = -1$, there are two potential
wells, one of a f\/inite depth on the left branch of function $P(v)$
and another inf\/initely deep and wide well but bounded from the
bottom on the right branch of function $P(v)$ (see
Fig.~\ref{f17}c).

\begin{figure}[t]
\centerline{\includegraphics[width=145mm]{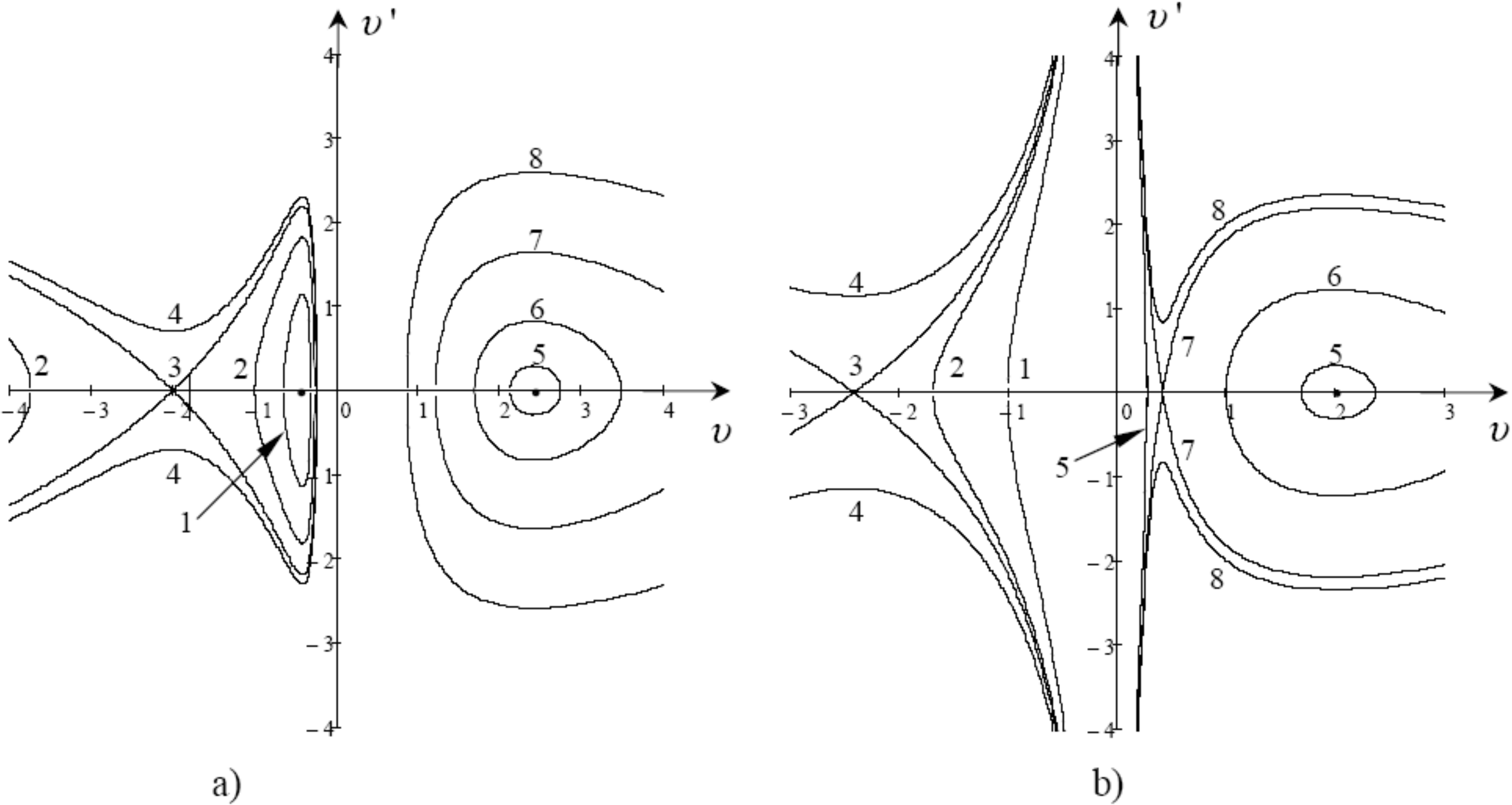}}

\caption{Phase plane corresponding to the subcritical case $C_1 <
C_1^{\rm cr}$. a) $C_1 = -5$, $C_2 = -1$. Line~1: $E = -6$; lines 2:
$E = -5$; line 3: $E = -4.25$; lines 4: $E = -4$; line 5: $E =
4.7$; line 6: $E = 5$; line~7: $E = 6$; line 8: $E = 8$. All
trajectories in the left half-plane outside of the closed loop of
separatrix correspond to unbounded solutions. b) $C_1 = -5$, $C_2
= 1$. Line 1: $E = -7$; line 2: $E = -5$; lines 3: $E = -4.657$;
lines 4: $E = -4$; lines 5: $E = 4.3$; line 6: $E = 5$; lines 7:
$E = 6.656$; lines 8: $E = 7$.}
\label{f19}%
\end{figure}

\begin{figure}[t]
\centerline{\includegraphics[width=145mm]{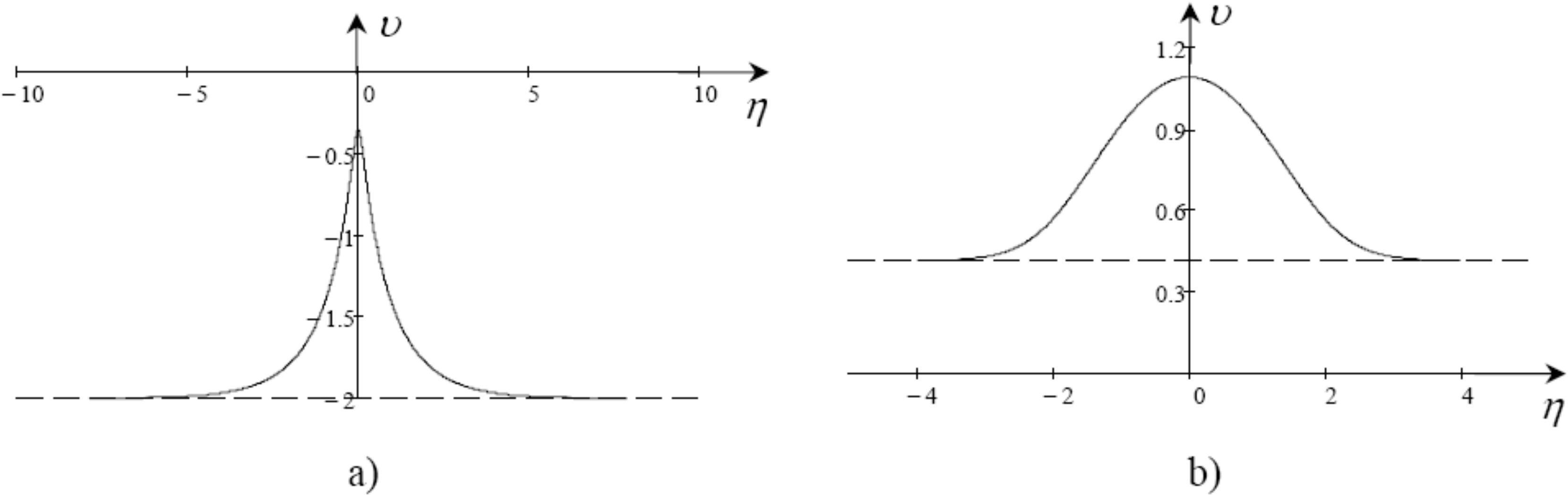}}

\caption{Soliton solutions on pedestals as described by
equations~\eqref{eq25}.}
\label{f20}%
\end{figure}

For the f\/irst potential well there is a family of closed
trajectories in the phase plane corresponding to periodic
solutions with the parameter $E$ varying between the local minimum
and maximum of the potential function; these solutions are
described by elliptic functions. All closed trajectories are
bounded by the loop of separatrix designated by symbol 3 in
Fig.~\ref{f19}a. Trajectories inside the separatrix loop next to
center correspond to quasi-sinusoidal waves, and the loop of the
separatrix corresponds to the solitary wave (soliton) which can be
treated as the limiting case of periodic waves. The soliton shape
is described by the following implicit formula:
\begin{gather}
\eta = \pm\sqrt{2}\left[\frac{v_1}{2\sqrt{v_2 - v_1}}
\ln{\left(\frac{\sqrt{v_2 - v_1} + \sqrt{v_2 - v}}{\sqrt{v_2 -
v_1} - \sqrt{v_2 - v}}\right)} + \sqrt{v_2 - v}\right], \qquad v_1
\le v \le v_2, \label{eq25}
\end{gather}
where $v_{1,2} = -\big(C_1 \mp \sqrt{C_1^2 - 3EC_2}\,\big)/E$,
$(v_1 < v_2)$ and $E = P_{\max}(C_1,C_2)$, where $P_{\max}(C_1,C_2)$
is the value of the potential local maximum shown in the left
half-plane of Fig.~\ref{f17}c. Solution \eqref{eq25} is shown in
Fig.~\ref{f20}a.

In original variables function $u$ describing soliton varies in
the range
\begin{gather*}
V + \frac{6v_1}{p + q} \le u \le V + \frac{6v_2}{p + q};
\end{gather*}
thus, the soliton amplitude amounts
\begin{gather}
A = 6\frac{v_2 - v_1}{p + q} = \frac{12}{p + q}\frac{\sqrt{C_1^2 -
3C_2P_{\max}(C_1,C_2)}}{P_{\max}(C_1,C_2)}; \label{eq27}
\end{gather}
The soliton velocity is
\begin{gather}
V = -\frac 1p\left[\beta + 2P_{\max}(C_1,C_2)\right]. \label{eq28}
\end{gather}

Equations \eqref{eq27} and \eqref{eq28} allow one to obtain a
direct relationship between the soliton's velocity and amplitude:
\begin{gather*}
A = -\frac{24}{p + q}\frac{\sqrt{C_1^2 + \frac 32 C_2(pV +
\beta)}}{pV + \beta}. 
\end{gather*}

For the second potential well located in the right half-plane of
Fig.~\ref{f17}c, there is another family of closed trajectories in
the phase plane corresponding to periodic solutions with the
parame\-ter~$E$ varying between the local minimum of the potential
function and inf\/inity; these trajectories are shown in the right
half-plane of Fig.~\ref{f19}a.

In the case of $C_2 = 1$, there is a shallow potential well on the
right branch of function~$P(v)$ and one inf\/initely deep well at
the origin where the potential function is singular. For the
shallow well there is a family of closed trajectories in the phase
plane corresponding to periodic solutions with the parameter $E$
varying between the local minimum and maximum of the potential
function. All such trajectories are also bounded by the loop of
separatrix designated by symbol~7 in Fig.~\ref{f19}b. The loop of
separatrix corresponds to the solitary wave whose shape is
described by the same implicit formula \eqref{eq25}, but with
dif\/ferent values of constants $C_1$, $C_2$, $E$ and
$P_{\max}(C_1,C_2)$, where $P_{\max}(C_1,C_2)$ is the value of the
potential local maximum shown in the right half-plane of
Fig.~\ref{f17}c. This solution is shown in Fig.~\ref{f20}b. All
above relationships between soliton amplitude and velocity, as
well as between soliton amplitude or velocity and constants $C_1$
and $C_2$ remain the same as above.

For the inf\/initely deep well at the origin there are two families
of compactons with nonpositive and nonnegative values; the phase
plane for them is similar to that shown in Fig.~\ref{f08} and
solutions are similar to those shown in Fig.~\ref{f09}. The entire
phase portrait of the system in the case of $C_2 = 1$ is shown in
Fig.~\ref{f19}b. Phase trajectories corresponding to positive
compactons are not shown in detail in that f\/igure because they are
too close to each other and are in the narrow gap between the axis
$v'$ and external two unclosed branches of the separatix~7 (only
one such trajectory, line~5, is shown in Fig.~\ref{f19}b; all
other trajectories are similar).

\section{Conclusion}
\label{sec5} %
As was shown in the paper, the extended reduced Ostrovsky equation
\eqref{eq1} possesses periodic and solitary type solutions in general.
There is a variety of solitary-wave solutions including compactons
with inf\/inite derivatives at the edges, smooth solitons, and
periodic waves. All compactons, however, actually are of the
compound-type solutions, i.e., they consist of two or more
non-smooth branches. Among periodic waves depending on the
equation parameters, there are also both smooth solutions and
compound-type solutions which consist of periodic sequences of
non-smooth branches (see, e.g., Fig.~\ref{f05}b). Moreover,
using compacton solutions as the elementary blocks, one can
construct very complex compound solutions including stochastic
stationary waves.

The approach used in this paper and based on the qualitative
theory of dynamical systems is free from the limitations of paper
\cite{Parkes} and allows us to present a complete classif\/ication
of all possible solutions of stationary exROE. In particular,
solutions were obtained and analyzed in details for the case $p +
q = 0$ that was out of consideration in paper~\cite{Parkes}.
Another ``prohi\-bi\-ted'' combination of parameters, $qV - \beta \ne
0$, that was also out of consideration in paper \cite{Parkes},
does not even appear in our study. The approach exploited in the
present paper is based on a~vivid mechanical analogy between a~particle moving in a special potential f\/ield and considered
stationary exROE. This approach allows one to observe
qualitatively an entire family of all possible solutions even
without construction of particular exact solution. A similar
approach has been exploited recently in application to the reduced
Ostrovsky equation~\cite{Stepanyants,Li} and exROE~\cite{Li},
although in the last case, the complete solution classif\/ication
was not considered.

\pdfbookmark[1]{References}{ref}

\LastPageEnding


\begin{thebibliography}{99}

\footnotesize\itemsep=0pt

\bibitem{Parkes} Parkes E.J., Periodic and solitary travelling-wave solutions of an
extended reduced Ostrovsky equation, {\it SIGMA} {\bf 4} (2008),
053, 17~pages, \href{http://arxiv.org/abs/0806.3155}{arXiv:0806.3155}.

\bibitem{Morrison-Parkes}
Morrison A.J., Parkes E.J.,
The $N$-soliton solution of the modif\/ied generalised Vakhnenko equation
(a new nonlinear evolution equation),
{\it Chaos Solitons Fractals} {\bf 16} (2003), 13--26.

\bibitem{Li}
Li J.-B.,
Dynamical understanding of loop soliton solution for several nonlinear wave equations,
{\it Sci. China Ser. A} {\bf 50} (2007), 773--785.


\bibitem{Stepanyants}
Stepanyants Y.A., On stationary solutions of the reduced Ostrovsky
equation: Periodic waves, compactons and compound solitons,
{\it Chaos Solitons Fractals} {\bf 28} (2006), 193--204.

\bibitem{Ostrovsky}
Ostrovsky~L.A., Nonlinear internal waves in a rotating ocean, {\it
Okeanologiya} {\bf 18} (1978), 181--191 (Engl.
transl: {\it Oceanology} {\bf 18} (1978), 119--125).

\end{thebibliography}
\end{document}